\documentclass[fleqn,usenatbib]{mnras}

\usepackage{newtxtext,newtxmath}
\usepackage[T1]{fontenc}

\DeclareRobustCommand{\VAN}[3]{#2}
\let\VANthebibliography\thebibliography
\def\thebibliography{\DeclareRobustCommand{\VAN}[3]{##3}\VANthebibliography}

\usepackage{graphicx}	% Including figure files
\usepackage{amsmath}	% Advanced maths commands
\usepackage{siunitx}
\usepackage{multirow}
\usepackage{bigdelim}
\usepackage{xcolor}

\newcommand{\citest}[2]{(#1; \textbf{\ref{obj:detection}}\citealt{#2})}

\newcommand{\smallcaps}[1]{{\small #1}}
\newcommand{\per}[1]{\mathrm{#1}^{-1}}
\newcommand{\cgslum}{\mathrm{erg}\ \per{s}}
\newcommand{\cgsflux}{\mathrm{erg}/\mathrm{cm}^2/\mathrm{s}}
\newcommand{\keV}{\mathrm{keV}}
\newcommand{\errs}[2]{^{+#1}_{-#2}}
\newcommand{\uptri}{\blacktriangle}
\newcommand{\dntri}{\blacktriangledown}

\newcommand{\solarM}{\mathrm{M}_{\odot}}
\newcommand{\Tin}{k T_\mathrm{in}}
\newcommand{\xmm}{{\it XMM-Newton} }
\newcommand{\nustar}{{\it NuSTAR} }
\newcommand{\swift}{{\it Swift} }
\newcommand{\chandra}{{\it Chandra} }

\newcommand{\hstacs}{{\it HST}/ACS }
\newcommand{\redchisq}{\chi^2_{\mathrm{red}}}

\title[IC5052 ULX: X-ray Analysis and Optical Search]{A Broadband X-ray Analysis and Optical Counterpart Search of IC5052 ULX}

\author[N. Brice et al.]{
Nabil Brice,$^{1}$\thanks{E-mail: nabil.brice2@herts.ac.uk}
Dominic J. Walton,$^{1}$
Felix F\"{u}rst$^{2}$,
Daniel Stern$^{3}$,
Amy H. Knight$^{4}$,
Matteo Bachetti$^{5}$,\newauthor
Matthew J. Middleton$^{6}$,
Alistair Pagan$^{4}$,
Tim P. Roberts$^{4}$,
Murray Brightman$^{7}$,
Hannah P. Earnshaw$^{7}$,\newauthor
Fiona A. Harrison$^{7}$,
Sean N. Pike$^{8}$
\\
$^{1}$Centre for Astrophysics Research, University of Hertfordshire, College Lane, Hatfield AL10 9AB, UK\\
$^{2}$European Space Astronomy Centre (ESA/ESAC), Operations Department, Villaneuva de la Ca\~{n}ada, 28692 Madrid, Spain\\
$^{3}$Jet Propulsion Laboratory, California Institute of Technology, 4800 Oak Grove Drive, Pasadena, CA 91109, USA\\
$^{4}$Centre for Extragalactic Astronomy, Department of Physics, Durham University, South Road, Durham DH1 3LE, UK\\
$^{5}$INAF-Osservatorio Astronomico di Cagliari, via della Scienza 5I-09047 Selargius (CA), Italy\\
$^{6}$Department of Physics and Astronomy, University of Southampton, Highfield, Southampton SO17 1BJ, UK\\
$^{7}$Cahill Center for Astrophysics, California Institute of Technology, 1216 East California Boulevard, Pasadena, CA 91125, USA\\
$^{8}$Department of Astronomy and Astrophysics, University of California, San Diego, CA 92093, USA\\
}

\date{Accepted XXX. Received YYY; in original form ZZZ}

\pubyear{2015}

\begin{document}
\label{firstpage}
\pagerange{\pageref{firstpage}--\pageref{lastpage}}
\maketitle

\begin{abstract}
We present broadband X-ray spectral and timing analysis of the Ultra-luminous X-ray source (ULX) in IC5052 using simultaneous \xmm and \nustar observations from 2022, supplemented by archival 2013 \xmm data. A two-thermal component model, often interpreted as radially-segregated emission from a super-Eddington inner disc and its associated wind, provides a statistically acceptable fit but yields an implausibly high inner disc temperature of $\Tin \approx 6.4~\keV$, inconsistent with even super-Eddington disc models. Including an additional continuum component from either an accretion column or a Comptonizing corona, as motivated by high S/N observations from other ULXs, provides comparable goodness of fit while allowing plausible inner disc temperatures. The accretion column model yields $\Tin \sim 1.2~\keV$ with the column contributing $F_\mathrm{col} \approx 62\%$ of total flux, while the Comptonizing corona model yields $\Tin \sim 3.0~\keV$ with a scattered fraction $\sim 1$, assuming the hotter disc provides the seed photons. Timing analysis initially challenges both scenarios: the accretion column model places IC5052 ULX where prior results suggest pulsations may be detectable ($F_\mathrm{col} \sim 60\%$), yet none were detected, while the corona model appears inconsistent with its lack of observed short-timescale variability. However, incorporating spectral information relaxes these constraints, allowing both models to remain physically plausible for IC5052 ULX. Finally, using improved \chandra astrometry, we identified a candidate optical counterpart consistent with an evolved high mass donor. A discrepancy between the optical extinction and X-ray fitted absorption suggests localised X-ray absorption.
\end{abstract}


%%%%%%%%%%%%%%%%%%%%%%%%%%%%%%%%%%%%%%%%%%%%%%%%%%

%%%%%%%%%%%%%%%%% BODY OF PAPER %%%%%%%%%%%%%%%%%%

\section{Introduction}
Ultra-luminous X-ray sources (ULXs) are extragalactic off-nuclear point sources, defined by an apparent X-ray luminosity exceeding $10^{39}~\cgslum$, which is the Eddington limit for $10~\solarM$ black holes. ULXs are understood to (mostly) represent a super-Eddington accretion state onto stellar-mass compact objects (see \citealt{king2023, pinto2023} for recent reviews), distinct from the sub-Eddington accretion modes typically observed in Galactic X-ray binaries. However, fundamental questions remain: the exact nature of the super-Eddington accretion state, the proportion of neutron star versus black hole accretors in the ULX population, and the physical mechanisms governing their spectral components, variability, outflows, and epoch-varying pulsed-fraction behaviour (for those with detected pulsations).

The discovery of pulsations in M82 X-2 \citep{bachetti2014}, and subsequently other ULXs \citep{furst2016, israel2017, carpano2018, sathyaprakash2019, rodriguezcastillo2020, pintore2025, ducci2025}, definitively established that neutron stars are capable of sustaining apparent luminosities up to several hundred times their Eddington limit. Currently, aside from dynamical mass measurements, the detection of coherent pulsations represents the only unambiguous method for identifying neutron star accretors. However, pulsation searches face significant observational challenges due to the low photon count rates and other complicating factors, such as the (unknown) orbital motion of the compact object and the large accretion-driven spin-up possibly shifting the pulse frequency during an observation. In the majority of known ULX pulsars (ULXPs), the pulsations are transient with duty cycles below $50\%$ \citep{rodriguezcastillo2020, bachetti2020}, without an understood physical model for this behaviour. The overall difficulty of direct pulsation detection has motivated alternative methods for identification of the compact accretor type.

Across a broadband X-ray ULX sample, \cite{walton2018} demonstrated the need for an additional high-energy spectral component beyond the two-component spectral models - previously established for the data below $10~\keV$ - in order to capture the hard excess above $\sim 10~\keV$. For ULXPs with sufficiently good quality broadband data, phase-resolved spectroscopy has demonstrated a relation between the hard excess and the pulsed emission \citep{brightman2016, walton2018a, walton2018}, which in accreting X-ray pulsars are thought to originate from the magnetically-confined accretion column structure that rises above the neutron star surface at sufficiently high accretion rates (see \citealt{mushtukov2023a} for a review; \citealt{basko1976} for the original model; \citealt{mushtukov2015} for its application to ULXs with strong dipole magnetic fields; \citealt{brice2021} for its application to ULXs with multi-polar magnetic field configurations). The spectral shape of the pulsed hard excess is well described with a cutoff power-law, giving a magnetic accretor spectral model consisting of two thermal components attributed to the super-Eddington disc and wind emission (double-thermal) plus a cutoff power-law representing the accretion column.

Remarkably, the same three-component spectral model successfully describes both confirmed ULXPs and non-pulsing ULXs, suggesting either a common emission mechanism or hinting at the possibility of a common accretor type across much of the ULX population. The success of fitting spectral models commonly used for Galactic X-ray pulsars to the $0.3-10~\keV$ ULX spectra is also suggestive of a neutron star accretor \citep{pintore2017, koliopanos2017}. From the sample of ULXs observed in the broadband that was available at the time (limited in number but including both the known ULXPs and non-pulsating sources), \cite{walton2018} noted an empirical trend when fitting with the magnetic accretor model: sources with accretion column flux fractions $F_\mathrm{col} \gtrsim 60\%$ measured over $0.3-40~\keV$ show pulsations, whereas those with $F_\mathrm{col} < 60\%$ typically do not. This is in line with the suggestion by \cite{king2017} that, for any particular ULX, the balance between the coherently pulsed accretion column emission and the disc emission determines the detectability of the pulsations.

An alternative spectral model consisting of a Comptonization up-scattering component applied to the hotter thermal component was also considered by \cite{walton2018}. This non-magnetic accretor spectral model is more similar in physical picture to the spectral models of sub-Eddington accreting black holes, where seed photons from the inner accretion disc are up-scattered by a hot corona. In Galactic X-ray binary systems, the coronae typically show broadband aperiodic variability arising from propagating mass accretion rate fluctuations in the disc \citep{churazov2001, demarco2023}. However, \cite{heil2009} found suppressed variability in a systematic study of ULXs, primarily constraining the $\lesssim 10~\keV$ band where the soft thermal components dominate. A Comptonizing corona would contribute more substantially to the spectrum above $\sim 10~\keV$, where the short-term variability could potentially provide a diagnostic to discriminate between these physical scenarios. Some tentative evidence for this was reported for Holmberg IX X-1 \citep{walton2014}, although the low count rates at higher energies present observational challenges.

Prior spectral classification of ULXs based solely on the $0.3-10~\keV$ data distinguished between three spectral states \citep{sutton2013}: "soft ultra-luminous" (SUL; dominated by soft thermal emission peaking below ${\sim}1~\keV$), "hard ultra-luminous" (HUL; characterised by harder emission with significant flux extending to higher energies), and "broadened disc" (BD; exhibiting a broad thermal component). These classifications were interpreted within a unified model in which the SUL and HUL states arose from different viewing geometries relative to the accretion disc and outflow \citep{poutanen2007, middleton2015}, while the BD state was attributed to differing accretion rates. However, the systematic requirement for an additional hard component in the broadband observation raises questions about whether these classifications and their physical interpretations remain valid, or whether they are artefacts of a more limited energy range.

IC5052 ULX, located in an edge-on galaxy at a distance of $5.5~\mathrm{Mpc}$ \citep{tully2013}, provides an interesting test case for these competing interpretations. This source was also recently studied by \citet{cruz-sanchez2025}, who analysed the 2022 broadband X-ray spectrum with both magnetic and non-magnetic accretor phenomenological models, before adopting the non-magnetic (super-critically accreting) black hole scenario for their detailed physical interpretation. In this work, we present an independent, more in-depth analysis of the same broadband dataset, and demonstrate the need from physical plausibility for an additional high-energy spectral component similar to other ULXs with broadband X-ray coverage \citep{walton2018}. We then examine the physical plausibility of both magnetic and non-magnetic accretor interpretations when accounting for the timing constraints. Finally, we search for an optical counterpart, considering the X-ray properties of the source in the context of the possible donor system. The paper is organised as follows: observations and data reduction are described in Section~\ref{sec:observations}; spectral analysis is presented in Section~\ref{sec:spectroscopy}; timing analysis, including variability characterisation and pulsation searches, in Section~\ref{sec:timing}; optical counterpart identification in Section~\ref{sec:optical}; discussion of the results in Section~\ref{sec:discussion}; and finally we give our conclusions in Section~\ref{sec:conclusions}.

\section{Observations and Data Reduction}
\label{sec:observations}
IC5052 ULX was simultaneously observed with \xmm \citep{jansen2001} and \nustar \citep{harrison2013} on 5th October 2022 (PI: Walton), enabling characterisation of the spectral components from $0.3-20~\keV$ (\S\ref{sec:spectroscopy}) and a search for pulsations (\S\ref{sec:timing}). We supplement this primary X-ray broadband dataset with earlier archival data, obtained from \xmm in 2013, {\it Chandra X-ray Observatory} ({\it Chandra}; \citealt{weisskopf2000}) in 2015, and intermittent monitoring from the {\it Neil Gehrels Swift Observatory} ({\it Swift}; \citealt{gehrels2004}) to examine long-term flux variations. Details of the observations used for the spectroscopic and timing analyses are given in Table~\ref{tab:observations}. The \swift XRT monitoring observations are detailed in \S~\ref{sec:timing}.

\begin{table}
  \begin{center}
  \caption{The X-ray observations of IC5052 ULX used in the spectroscopic analysis. For \textit{XMM-Newton}, net exposures are listed for Epn and combined EMOS detectors; for \textit{NuSTAR}, net exposures (combined from science mode 01 and spacecraft mode 06) are for FPMA and FPMB.}
  \label{tab:observations}
  \begin{tabular}{lccc}
    \hline
    \hline
    Observatory & Observation ID & Start Date & Exposure (ks)\\
    \hline
    \multicolumn{4}{c}{\textit{Primary Broadband Data}}\\
    \hline
    \xmm & 0912390101 & 2022-10-05 & 46.12, 47.98\\
    \nustar & 30801029002 & 2022-10-05 & 107.7, 106.6\\
    \hline
    \multicolumn{4}{c}{\textit{Archival Data}}\\
    \hline
    \xmm & 0721910501 & 2013-10-01 & 14.60, 17.95\\
    \chandra & 17000 & 2015-05-24 & 1.02\\
    \hline
  \end{tabular}
  \end{center}
\end{table}

\subsection{XMM-Newton} \label{sec:data-reduction-XMM-Newton}
We reduced the \xmm data using the \xmm Science Analysis Software (\smallcaps{SAS}) v21.0.0, largely following the standard procedure~\footnote{\url{https://www.cosmos.esa.int/web/xmm-newton/} \label{fn:sas-operations-webpage}}. Raw observation data files were processed using the \texttt{epproc} and \texttt{emproc} tasks to produce the calibrated event lists for the EPIC-pn (Epn; \citealt{struder2001}) and EPIC-MOS (EMOS; \citealt{turner2001}) detectors respectively. As there was heightened background activity across a significant portion of the 2022 observation, where the Epn count rate exceeded $0.4 \ \text{ct s}^{-1}$ in the $10-12~\keV$ energy band, we used the \texttt{maxSNR} procedure outlined by \cite{piconcelli2004} to determine good-time-intervals that maximise the S/N in the $0.35-8~\keV$ energy band - this narrower energy band than the standard ($0.3-10~\keV$) \xmm energy band was chosen after a first pass in spectral binning for S/N $>5$. The resulting Epn total exposure time increased compared to the standard background filtering procedure, from ${\sim}21$ ks to ${\sim}46$ ks. In contrast, the EMOS total exposure time was not significantly increased (from $47.97~\text{ks}$ to $47.98~\text{ks}$), as the particle background flaring primarily affects the Epn detector\footnote{XMM-Newton Users Handbook, Issue 2.23 (ESA: XMM-Newton SOC, 2025), \url{https://xmm-tools.cosmos.esa.int/external/xmm_user_support/documentation/uhb/}}. We found the spectral shape to be in good agreement with the one obtained using standard filtering procedures. No such particle background flaring significantly affected the 2013 observation, so in this case we generated the good-time-intervals for Epn and EMOS by applying the recommended count-rate cutoffs (for events above $10 \ \keV$) of $0.35 \ \text{ct s}^{-1}$ and $0.4 \ \text{ct s}^{-1}$ respectively. Finally, to enable accurate timing analysis, times were converted to the (solar system) Barycentric Dynamical Time (TBD) using \texttt{barycen} (JPL DE-200 ephemeris).

The source products were extracted from a ${\sim}26\arcsec$ radius circular region, which we selected to avoid the nearby chip gap in the Epn dataset. The background spectrum was estimated from a larger circular, source-free region on the same CCD. The cleaned and background-filtered Epn/EMOS event lists were filtered for up to double pixel pattern events (\texttt{PATTERN<=4}) / up to quadruple pixel pattern events (\texttt{PATTERN<=12}), with the appropriate good-events flag filter for each detector. The instrument responses for each of the detectors were generated from the \smallcaps{SAS} routines \texttt{rmfgen} and \texttt{arfgen}. Finally, after checking their consistency, the spectra from the EMOS1 and EMOS2 detectors were combined using the \smallcaps{SAS} routine \texttt{epicspeccombine}.

\subsection{NuSTAR}
We reduced and calibrated the \nustar data using the \nustar Data Analysis Software v2.1.4 and \nustar CALDB v20240325, following the recommended procedures. \texttt{nupipeline} was first used to generate the clean and filtered event files. We then extracted the source and instrument response products from a $40\arcsec$ radius (chosen to maximise the S/N) circular region for both FPMA/B with \texttt{nuproducts}, and the background was estimated from a larger $110\arcsec$ radius region on the same detector. To improve the SNR, we also extracted the `spacecraft science' data following the procedure outlined by \cite{walton2016}. Only those data with $\mathrm{S/N} > 5$ (which we estimated from the counts in the source and background regions) were combined into the final dataset. This provided ${\sim}10\%$ of the final good exposure. The spectra and instrument response files from the `spacecraft science' and `science' modes for each of FPMA and FPMB were combined using \texttt{addspec}.

\subsection{Chandra}
We reduced the \chandra observation using CIAO v4.17 and CALDB v4.12.0 following the standard procedures for ACIS data reprocessing\footnote{\url{https://cxc.cfa.harvard.edu/ciao/guides/acis_data.html}}. Source and background spectra were extracted using \texttt{specextract}, which also generated the corresponding response files. We extracted the source counts from a $\sim 4\arcsec$ radius circular region, and estimated the background from a $\sim 5\arcsec-20\arcsec$ annulus. The astrometric solution from the \textit{Chandra} Source Catalog (CSC) v2.1 \citep{Evans2024}, registered to Gaia eDR3 \citep{gaia-collaboration2021}, provided the positional information used for optical counterpart identification (\S\ref{sec:optical}).

\section{Spectroscopy}
\label{sec:spectroscopy}
For the spectral analysis of all observations, we rebinned the events using \texttt{ftgrouppha} to a minimum signal-to-noise ratio (SNR) of 5 and at least 25 counts, enabling analysis with $\chi^2$ statistics. This binning restricted the energy range of the broadband observations to $0.35 - 20 \ \keV$ ($0.35-8~\keV$ for {\it XMM-Newton}; $3-20~\keV$ for {\it NuSTAR}). Model fitting was performed using the \texttt{Sherpa} package (CIAO v4.17) with spectral models from XSPEC v12.14.0 \citep{arnaud1996}. We used a two-step optimisation procedure: first fitting with the Levenberg-Marquardt algorithm, followed by the \texttt{simplex} method in \texttt{Sherpa} to ensure a robust convergence.

For all spectral models, we performed simultaneous fits to the instrument-specific datasets. To account for cross-instrument calibration differences, we applied a constant normalisation factor, fixing Epn factor to unity while allowing the EMOS and \nustar FPM factors to vary freely. The resulting normalisation factors differed by $<10\%$ from Epn in the models with a sufficiently good fit, consistent with expectations \citep{madsen2015}. We included two absorption components \citest{\texttt{TBabs}}{wilms2000} in all models, one fixed at the Galactic column density along the line-of-sight to the source ($N_{\rm H} = 4.05 \times 10^{20} \ \mathrm{atoms / cm}^{2}$ from the HEASARC nH tool\footnote{\url{https://heasarc.gsfc.nasa.gov/cgi-bin/Tools/w3nh/w3nh.pl}}; \citealt{hi4pi-collaboration2016}), and a second absorption component with a free column density to model the intrinsic absorption related to the source and its local environment.

\begin{figure}
  \includegraphics[width=\columnwidth]{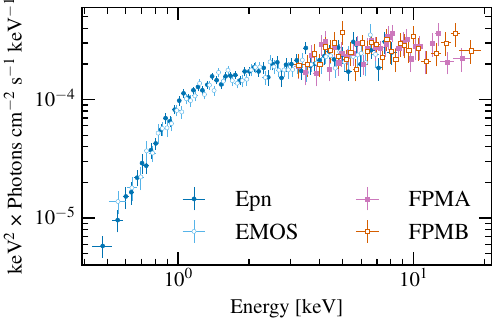}
  \caption{The broadband X-ray spectrum of IC5052 ULX from October 2022.
    The data from \xmm (Epn and EMOS) and \nustar (FPMA and FPMB) are unfolded with a powerlaw model of photon index $\Gamma = 0$ and logarithmically rebinned to 50 bins per energy decade for visual clarity.}
  \label{fig:ufspec}
\end{figure}

\begin{figure}
  \includegraphics[width=\columnwidth]{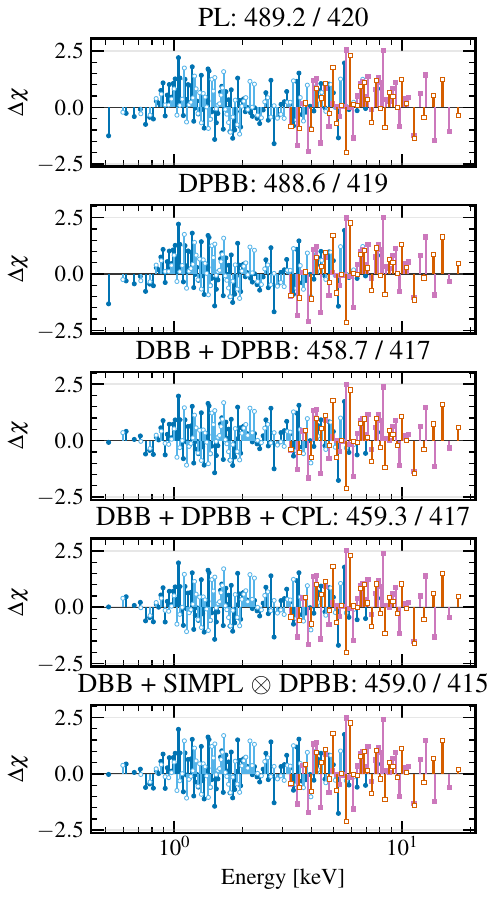}
  \caption{The \xmm (Epn and EMOS shown by filled and hollow circular markers, respectively) and \nustar (FPMA and FPMB shown by filled and hollow square markers, respectively) data residuals, calculated as $(\mathrm{Data} - \mathrm{Model}) / \mathrm{Error}$. From top to bottom, the panels show residuals for an absorbed powerlaw model (PL), thermal model (DPBB), double thermal model (DBB + DPBB), magnetic accretor model (DBB + DPBB + CPL), and non-magnetic accretor model (DBB + SIMPL $\otimes$ DPBB), respectively. The Epn and EMOS are rebinned by a factor of $2$ for clarity. The $\chi^2$~/~d.o.f. is shown alongside the model name.}
  \label{fig:ratios}
\end{figure}

The unfolded X-ray broadband spectrum of IC5052 ULX is shown in Figure~\ref{fig:ufspec}. The spectral rollover ubiquitously seen in \nustar observations of other ULXs - Circinus ULX5 \citep{walton2013}, NGC1313 X-1 \citep{bachetti2013}, Holmberg II X-1 \citep{walton2015}, IC342 X-1 and X-2 \citep{rana2015}, NGC5204 X-1 \citep{mukherjee2015}, Holmberg IX X-1 \citep{walton2015}, NGC5907 ULX \citep{furst2017} - is present here too, though appears to be much more subtle than in the majority of these other cases. Despite this relatively mild curvature, an absorbed powerlaw model (PL; best-fit photon index of $\Gamma \sim 1.93$) still provides a poor fit to the 2022 broadband data, yielding $\redchisq \sim 489.2 / 420$ (null hypothesis probability of $0.011$). The residuals for this best-fit PL model (top panel of Figure~\ref{fig:ratios}) reveals an excess curvature in both the $0.4 - 2 \ \keV$ and $5 - 25 \ \keV$ bands. Even restricting the fit to only the \xmm data provides a relatively poor fit ($\redchisq \sim 411.8 / 352$, null hypothesis probability of $0.015$), suggesting that this simple, single-component model still does not sufficiently describe the shape of the soft X-ray data. In contrast, the shorter-exposure 2013 archival \xmm data is adequately fit by a similar absorbed powerlaw model, with $\redchisq \sim 171.6 / 186$, although this is likely due to the weaker statistical power rather than genuine differences in spectral complexity. In fact, the similar photon index ($\Gamma \approx 1.94$) in the best-fit and comparable luminosity (see \S\ref{sec:timing}) suggests the underlying spectral state is consistent with that during the 2022 observations.

Since ULXs are expected to be powered by accretion onto a compact object, we next tested the fit of thermal models, starting with a single thermal-component model: \texttt{diskpbb} (DPBB; \citealt{mineshige1994}). This model includes a free parameter $p$ for the radial temperature profile ($T \propto r^{-p}$, where $T$ is the local disc temperature, and $r$ is the radial distance from the accretor), generalising the standard \cite{shakura1973} thin accretion disc, for which $p = 0.75$. The best-fit model has $p \approx 0.5$, similar to the slim-disc model \citep{abramowicz1988} expected for super-Eddington accretion flows. The full set of best-fit parameters for this model are reported in Table~\ref{tab:best-fit-model}. However, as with the absorbed powerlaw model, the overall fit to the 2022 broadband data was poor ($\redchisq \sim 488.6 / 419$) and unphysical (as $\Tin \gtrsim 8.7~\keV$; Table~\ref{tab:best-fit-model}), motivating model fits with additional spectral components.

The poor fit of the single thermal-component model can be improved by introducing a second thermal component, an approach commonly required for high-quality ULX spectra in the $0.3 - 10 \ \keV$ range (see e.g. \citealt{stobbart2006}, \citealt{gladstone2009}). In this two-component spectral model of the ultra-luminous accretion state, the hotter thermal component is interpreted as emission from the inner regions of a super-critical accretion disc, while the cooler thermal component is attributed to either the outflowing wind or a standard sub-Eddington thin disc at larger radii. We found the double thermal-component model (\texttt{diskbb + diskpbb}) provides a statistically acceptable fit to the broadband spectrum of IC5052 ULX, yielding $\redchisq \sim 458.7 / 417$, which is an improvement of $\Delta \chi^2 \approx 29.9$ for 2 additional free parameters compared to the single-component DPBB model. This improvement is statistically significant at $>3\sigma$ confidence according to the F-test. The residuals show no systematic structure across the broadband energy range (see panel labelled ``DBB + DPBB'' in Figure~\ref{fig:ratios}), indicating the spectral shape is adequately captured by this model.

The full set of best-fit parameters are reported in Table~\ref{tab:best-fit-model}. However, despite the reasonably good statistical fit, the best-fit model parameters give unphysical results. The inner disc temperature is significantly higher than that found in other ULXs fitted with similar models, $\Tin \approx 6.4 \ \keV$ ($\gtrsim 5 \ \keV$ at $90\%$ confidence), and the normalisation gives an unphysically small apparent emitting radius (see \S\ref{sec:broadband-continuum-emission}). Together, these present a challenge for physical interpretation and motivate the exploration of models with additional (high energy) spectral components that may allow a more realistic inner disc temperature.

Finally, in order to place IC5052 ULX within the context of ULX spectral classification using only the $0.3 - 10~\keV$ spectra, we tested a fit of the two-component model (\texttt{diskbb + powerlaw}; DBB + PL) used by \cite{sutton2013} to establish their ultraluminous state classification scheme. We found this spectral model to adequately fit the \xmm-only dataset, yielding $\redchisq \sim 387.8 / 350$ (a statistically significant improvement over the PL model), but is unable to capture the high-energy rollover present in the \nustar dataset, yielding $\redchisq \sim 464.1 / 418$ (contrast with the DBB + DPBB model above). For the best-fit (when both including and excluding the \nustar dataset) model, we found $\Tin \approx 0.23\pm0.08$ and $\Gamma \approx 1.83\pm0.1$, which places IC5052 ULX within the Hard Ultra-Luminous (HUL) spectral category (since $\Gamma < 2$; \citealt{sutton2013}). This classification, and the underlying spectral parameters, are in agreement with the independent broadband analysis of \cite{cruz-sanchez2025}, who applied the same \citet{sutton2013} scheme to IC5052 and obtained a HUL class with $\Tin = 0.21\errs{0.05}{0.03}~\keV$ and $\Gamma = 1.82\pm0.01$.

\subsection{Magnetic Accretor Model}
\label{sec:magacc}
A pulsar-like spectral model (\texttt{diskbb + diskpbb + cutoffpl}; DBB + DPBB + CPL) has been successfully fitted to a sample of broadband ULXP spectra \citep{walton2018} when performed as part of a joint spectral-timing analysis, where the CPL photon index $\Gamma$ and the high-energy cutoff $\mathrm{HighECut}$ are constrained from phase-resolved spectral analyses across the pulsation periods. In this model, the CPL component is interpreted as emission from the neutron star accretion column \citep{pintore2017}. Due to the lack of detectable pulsations in our dataset (see \S\ref{sec:timing}), we fixed the CPL shape parameters to the average of values found by \cite{walton2020} for the ULXP sample: $\Gamma = 0.59$ and $\mathrm{HighECut} = 7.9 \ \keV$. After fitting with fixed parameters, we tested a fit allowing for free CPL shape parameters, but this provided negligible improvement to the statistic ($\Delta \chi^2 < 1$ for 2 fewer degrees of freedom), confirming that our spectroscopic data are not able to provide adequate constraints for these parameters.

\begin{figure*}
  \includegraphics[width=\textwidth]{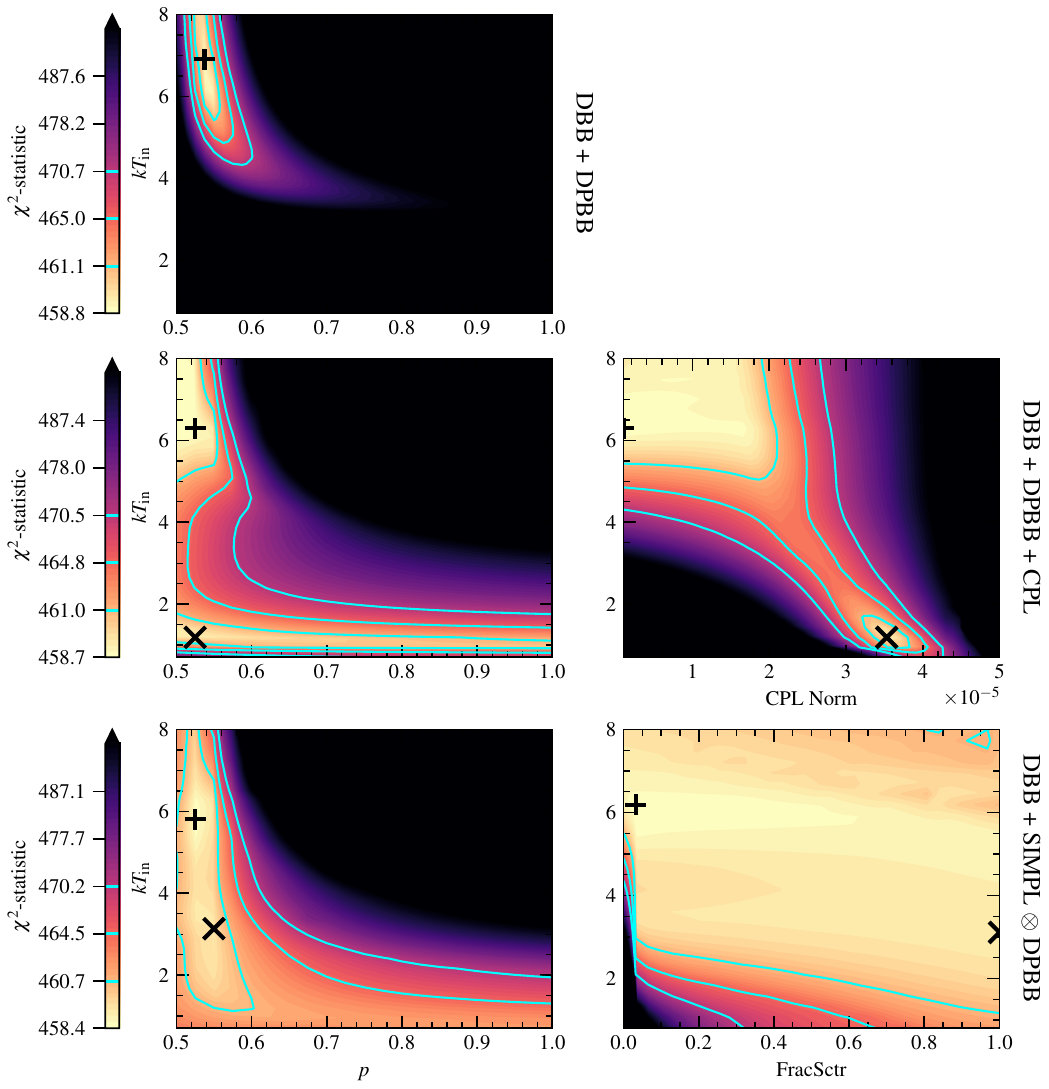}
  \caption{The $\chi^2$-surfaces across two model parameters, where the colours are scaled according to the minimum $\chi^2$ for each row and $6\sigma$ upper limit for 2 degrees of freedom. Each row corresponds to a different model (from top to bottom: DBB + DPBB, DBB + DPBB + CPL, DBB + SIMPL $\otimes$ DPBB). Panels of the left column show the $\chi^2$-surface across DPBB $T_\mathrm{in}$ and $p$ to illustrate how an additional spectral component (either CPL or SIMPL) modifies the $\chi^2$ landscape compared to the two-component DBB + DPBB model (top). Panels of the right column show the $\chi^2$-surface across DPBB $T_\mathrm{in}$ and the additional spectral component normalisation parameter (CPL $\mathrm{Norm}$ or SIMPL $\mathrm{FracSctr}$). The cyan contours mark the $1\sigma, \ 2\sigma, \ 3\sigma$ confidence levels for 2 degrees of freedom, with corresponding $\chi^2$ values shown in the associated colour bar. In each panel, the $\times$-marker indicates the location of the model local best-fit from Sherpa optimisation, and the $+$-marker indicates the location of the global best-fit.}
  \label{fig:contours}
\end{figure*}

\begin{figure}
  \includegraphics[width=\columnwidth]{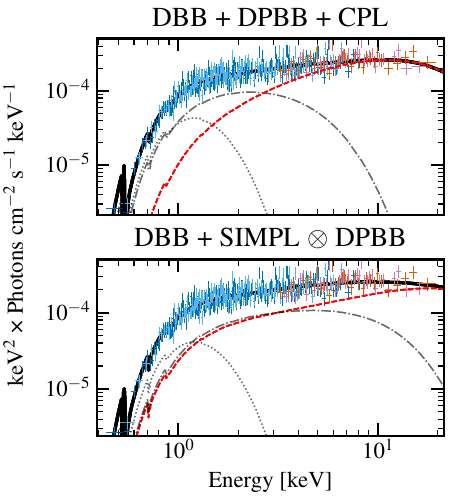}
  \caption{Unfolded broadband spectra showing the best-fit magnetic accretor model (top panel; DBB + DPBB + CPL) and non-magnetic accretor model (bottom panel; DBB + SIMPL $\otimes$ DPBB). In both panels, the solid line shows the total model spectrum. For the magnetic accretor model, the dotted, dash-dotted, and dashed lines show the DBB, DPBB, and CPL components respectively. For the non-magnetic accretor model, the dotted line shows the DBB component, the dash-dotted line shows the seed DPBB spectrum before Comptonization, and the dashed line shows the Comptonized emission (where the underlying seed spectrum has been removed from SIMPL $\otimes$ DPBB).}
  \label{fig:eeufspectra-bestfit}
\end{figure}

With the fixed CPL shape, we found two distinct regions of the parameter space with statistically good fits (see Figure~\ref{fig:contours}). The global minimum is located at $\Tin \approx 6.4~\keV$ with $\redchisq \sim 458.7 / 416$, where the parameters are nearly identical to the DBB + DPBB model and the CPL contribution to the broadband spectrum is negligible. However, for the best-fit in the second region located at $\Tin \approx 1.2~\keV$, the CPL normalisation is sufficiently high to meaningfully contribute to the broadband spectrum. This local best-fit (Table~\ref{tab:best-fit-model}) is of comparable quality to the global best-fit, with $\redchisq \sim 459.3/416$ and $\Delta \chi^2 \approx +0.6$ for the same number of degrees of freedom, which is within the range defining the $90\%$ confidence interval. The inner disc temperature likewise remains consistent with other broadband ULXs \citep{walton2018}. The total magnetic accretor model and component unfolded spectra for the local best-fit are shown in the top panel of Figure~\ref{fig:eeufspectra-bestfit}. The residuals show (in the panel labelled ``DBB + DPBB + CPL'' of Figure~\ref{fig:ratios}) no systematic structure across the broadband energy range.

We obtained the observed model flux estimates (in the $0.35 - 20~\keV$ broadband energy range) of the local best-fit by randomly sampling $10,000$ models within the $90\%$ confidence intervals for each parameter and computing the weighted 5th and 95th quantiles (using the $\chi^2$ likelihood) for an interval containing $90\%$ of the total weighted samples. In the extrapolated $0.3 - 40~\keV$ energy band, which we used for direct comparison with \cite{walton2018}, the CPL component was found to contribute $68.3^{+0.9}_{-3.2}\%$ of the emission.

\subsection{Non-Magnetic Accretor Model}
\label{sec:bhmodel}
An alternative spectral model applicable to ULXs with black hole accretors consists of a \texttt{simpl} component \citep{steiner2009} applied to the double thermal model (in particular convolved with the hotter thermal component: \texttt{diskbb + simpl}$\otimes$\texttt{diskpbb}; DBB + SIMPL $\otimes$ DPBB). In this model, the SIMPL component is interpreted as Comptonization of seed photons from the hot inner accretion disc (DPBB) by a corona, and is parameterised by a powerlaw photon index $\Gamma$ and a scattering fraction $\mathrm{FracSctr}$ of seed photons.

We found a local best-fit of the non-magnetic accretor model, where $\mathrm{FracSctr} > 0$ and $\Tin \approx 3.0$ - comparable to values seen in other ULXs, with $\redchisq \sim 459.0 / 415$ (see Table~\ref{tab:best-fit-model} for full details of other parameter values). However, similar to the magnetic accretor model, the global best-fit is located at $T_\mathrm{in} \approx 6.4~\keV$, where $\mathrm{FracSctr} \approx 0$ (meaning the SIMPL component is negligible) and with $\redchisq \sim 458.3 / 415$ (so the local best-fit is $\Delta \chi^2 \approx 0.7$ for the same number of degrees of freedom, again like the magnetic accretor model comfortably within the range for the $90\%$ confidence interval). Unlike the distinct local-minimum in the $\chi^2$-surface of the magnetic accretor model, Figure~\ref{fig:contours} shows a wide range of $\mathrm{FracSctr}$ and $\Tin$ within the $1\sigma$ uncertainty. The parameter space opens up rapidly to lower temperatures as $\mathrm{FracSctr}$ increases from $0$, and a local minimum emerges at $\Tin \approx 3.0~\keV$ for $\mathrm{FracSctr} \approx 1$. Notably, both (global and local) minima lie at the bounds for $\mathrm{FracSctr}$, rather than converging to well-constrained values within the allowed range, so the associated best-fit parameter values should be treated with caution. The bottom panel of Figure~\ref{fig:eeufspectra-bestfit} displays the total model and component unfolded spectra for this local best-fit. The residuals (presented in the panel labelled ``DBB + SIMPL$\otimes$DPBB'' in Figure~\ref{fig:ratios}) show no significant systematic structure across the broadband energy range.

Using the same approach as for the magnetic accretor model, we estimated the observed model flux (in the $0.35 - 20~\keV$ broadband energy range) with weighted quantiles. We also estimated the photon flux contribution of the SIMPL component in the \xmm energy band ($0.3-10~\keV$; relevant for the variability analysis in \S\ref{sec:variability-analysis}) for 10,000 samples by computing the fraction of the difference in photon flux between the model components before and after SIMPL convolution. For the local best-fit model, the SIMPL component contributes $41.7^{+2.0}_{-2.8}\%$ of the photon flux in the $0.3-10~\keV$ energy band.

\begin{table*}
  \def\arraystretch{1.3}
  \begin{center}
    \caption{Best-fit model parameters
      for the continuum emission of IC5052 ULX, given with $90\%$ confidence intervals.}
  \label{tab:best-fit-model}
  \begin{tabular}{@{}r@{\hspace{1mm}}l@{\hspace{6pt}}l@{\hspace{8pt}}cccc@{}}
    \cline{3-7}
    & & Parameter & DPBB & DBB + DPBB & DBB + DPBB + CPL & DBB + SIMPL $\otimes$ DPBB \\
    \cline{3-7}
    & & $\chi^2$ / d.o.f. & 488.6 / 419 & 458.7 / 417 & 459.3 / 416 & 459.0 / 415 \\
    \\[0.8em]
    \multirow{3}{*}{\texttt{factor}} & \ldelim\{{3}{1mm} & EMOS & $0.94^{+0.03}_{-0.03}$ & $0.94^{+0.03}_{-0.03}$ & $0.94^{+0.03}_{-0.03}$ & $0.94^{+0.03}_{-0.03}$ \\
     &  & FPMA & $1.12^{+0.08}_{-0.08}$ & $1.05^{+0.08}_{-0.07}$ & $1.05^{+0.08}_{-0.07}$ & $1.05^{+0.08}_{-0.07}$ \\
     &  & FPMB & $1.14^{+0.09}_{-0.08}$ & $1.07^{+0.09}_{-0.08}$ & $1.07^{+0.09}_{-0.08}$ & $1.07^{+0.08}_{-0.07}$ \\
    \\[0.8em]
    \texttt{ztbabs} & & \shortstack[l]{$N_\mathrm{H}$ \\{}[$10^{22}$ atoms/cm$^{2}$]} & $0.48^{+0.03}_{-0.02}$ & $0.58^{+0.10}_{-0.07}$ & $0.60^{+0.07}_{-0.13}$ & $0.59^{+0.02}_{-0.10}$ \\
    \\[0.8em]
    \multirow{2}{*}{\texttt{diskbb}} & \ldelim\{{2}{1mm} & $\Tin$ [keV] & - & $0.29^{+0.06}_{-0.07}$ & $0.27^{+0.11}_{-0.06}$ & $0.27^{+0.10}_{-0.02}$ \\
     &  & Norm & - & $2.5^{+10}_{-2}$ & $4.0^{+20}_{-3}$ & $3.4^{+1.6}_{-2.6}$ \\
    \\[0.8em]
    \multirow{3}{*}{\texttt{diskpbb}} & \ldelim\{{3}{1mm} & $\Tin$ [keV] & $8.69^{\uptri}_{-0.81}$ & $6.41^{+3.29}_{-1.06}$ & $1.21^{+0.74}_{-0.40}$ & $3.02^{\uptri}_{-0.41}$ \\
     &  & $p$ & $0.51^{+0.01}_{-0.01}$ & $0.54^{+0.02}_{-0.02}$ & $0.52^{\uptri}_{\dntri}$ & $0.54^{+0.2}_{\dntri}$ \\
     &  & Norm [$10^{-5}$] & $0.18^{+0.09}_{-0.09}$ & $0.90^{+1.15}_{-0.68}$ & $250^{+2820}_{-200}$ & $7.5^{+143.6}_{-3.2}$ \\
    \\[0.8em]
    \texttt{cutoffpl} & & Norm [$10^{-4}$] & - & - & $0.36^{+0.02}_{-0.04}$ & - \\
    \\[0.8em]
    \multirow{2}{*}{\texttt{simpl}} & \ldelim\{{2}{1mm} & $\Gamma$ & - & - & - & $2.46^{+0.55}_{\dntri}$ \\
     &  & FracSctr & - & - & - & $1.000^{\uptri}_{\dntri}$ \\
    \\[0.8em]
     \shortstack[r]{$0.35-20~\keV$\\Flux} & & \shortstack[c]{$[10^{-13}~\cgsflux]$} & $9.16^{+0.19}_{-0.20}$ & $9.50^{+0.25}_{-0.18}$ & $9.55^{+0.23}_{-0.30}$ & $9.49^{+0.31}_{-0.22}$ \\
    \\[0.8em]
    \shortstack[r]{$0.35-20~\keV$\\$L_\mathrm{absorbed}^\star$} & & \shortstack[l]{$[10^{39}~\cgslum]$} & $3.31^{+0.07}_{-0.07}$ & $3.44^{+0.09}_{-0.07}$ & $3.46^{+0.08}_{-0.11}$ & $3.43^{+0.11}_{-0.08}$ \\
    \\[0.8em]
    \shortstack[r]{$0.35-20~\keV$\\(unabsorbed) $L^\star$} & & \shortstack[l]{$[10^{39}~\cgslum]$} & $4.66^{+0.09}_{-0.09}$ & $5.16^{+0.26}_{-0.04}$ & $5.36^{+0.19}_{-0.01}$ & $5.24^{+0.36}_{-0.18}$ \\
    \\[0.8em]
    \cline{3-7}
  \end{tabular}
  \end{center}
  Note: The parameters for \texttt{cutoffpl} (CPL) component were fixed at the averaged values used for the broadband spectrum analysis by \cite{walton2018}: photon index $\Gamma=0.59$, and high energy cutoff $\mathrm{HighECut}=7.9 \ \keV$. The data do not provide adequate constraints to fit these as free-parameters. \\
$^\uptri,^\dntri$: The $90\%$ confidence intervals for these parameters run up against the top and bottom parameter bounds, respectively. \\
Confidence intervals for the flux were computed after fitting by sampling $10,000$ models around the best-fit (using the confidence intervals) and taking the weighted quantiles (using the $\chi^2$ likelihood) at $\leq 0.05$ and $\leq 0.95$ of the cumulative sample weight for the lower and upper bounds, respectively. This was done to avoid overly broad confidence regions from the non-gaussian tails. \\
$^\star$: Isotropic luminosities are calculated taking the smaller $5.5~\mathrm{Mpc}$ distance \citep{tully2013}, which was obtained using the Tip of the Red Giant Branch method. An alternative distance of $6.85~\mathrm{Mpc}$ \citep{tully2016}, which was obtained using the Tully-Fisher method, yields higher luminosities by a factor of $\sim 1.55$.
\end{table*}

\section{Timing Analysis}
\label{sec:timing}
IC5052 ULX was first observed in the X-ray band by {\it XMM-Newton} in 2013 and subsequently monitored intermittently by \swift XRT \citep{gehrels2004, burrows2005}. From June 2014 to August 2018, \swift XRT accumulated 18 observations (ObsIDs 00084544001--00084544021) in Photon Counting mode, with a total exposure of $\approx 14.8~\mathrm{ks}$ and a mean of $\sim 0.8~\mathrm{ks}$ per pointing. The shortest exposure lasted $\approx 0.2~\mathrm{ks}$, yielding $8$ photon counts from the source region, while the longest exposure lasted $\approx 2.5~\mathrm{ks}$, yielding $86$ photon counts from the source region. In 2015, the source was also observed by {\it Chandra}. However, only the 2013 and 2022 {\it XMM-Newton} observations offer sufficient exposure for a deeper timing (and spectroscopic) analysis, and in the case of the 2022 observation was designed (PI: Walton) to collect enough photons for a meaningful pulsation search.

In order to assess the long-term variability seen from IC5052 ULX, we compiled all the available X-ray observations. The \swift XRT count-rate light curve was generated using the online product builder\footnote{\url{https://www.swift.ac.uk/user_objects/}} \citep{evans2009}, adopting per-observation binning in the standard $0.3-10~\keV$ band. As many of the individual \swift XRT observations contain too few counts to constrain a spectral model, we adopted the \swift XRT count-rate as the metric for the long-term light curve, converting fluxes from \xmm and \chandra data using the \texttt{WebPIMMS} tool\footnote{\url{https://heasarc.gsfc.nasa.gov/cgi-bin/Tools/w3pimms/w3pimms.pl}}. Since the \texttt{WebPIMMS} tool only allows for simple spectral models (and in any case the \chandra dataset is not sufficiently constraining for a multi-component model), we modelled the continuum with an absorbed powerlaw model. The hydrogen column density was fixed to the value obtained from the 2022 \xmm dataset owing to the uncertainties from low exposure times in all other datasets. Similarly, the powerlaw index was fixed to the value obtained from the 2022 {\it XMM-Newton} dataset, which was also within $1\sigma$ uncertainty of the powerlaw index obtained from the 2013 \xmm dataset, before fitting the \chandra dataset.

\begin{figure*}
  \includegraphics[width=\textwidth]{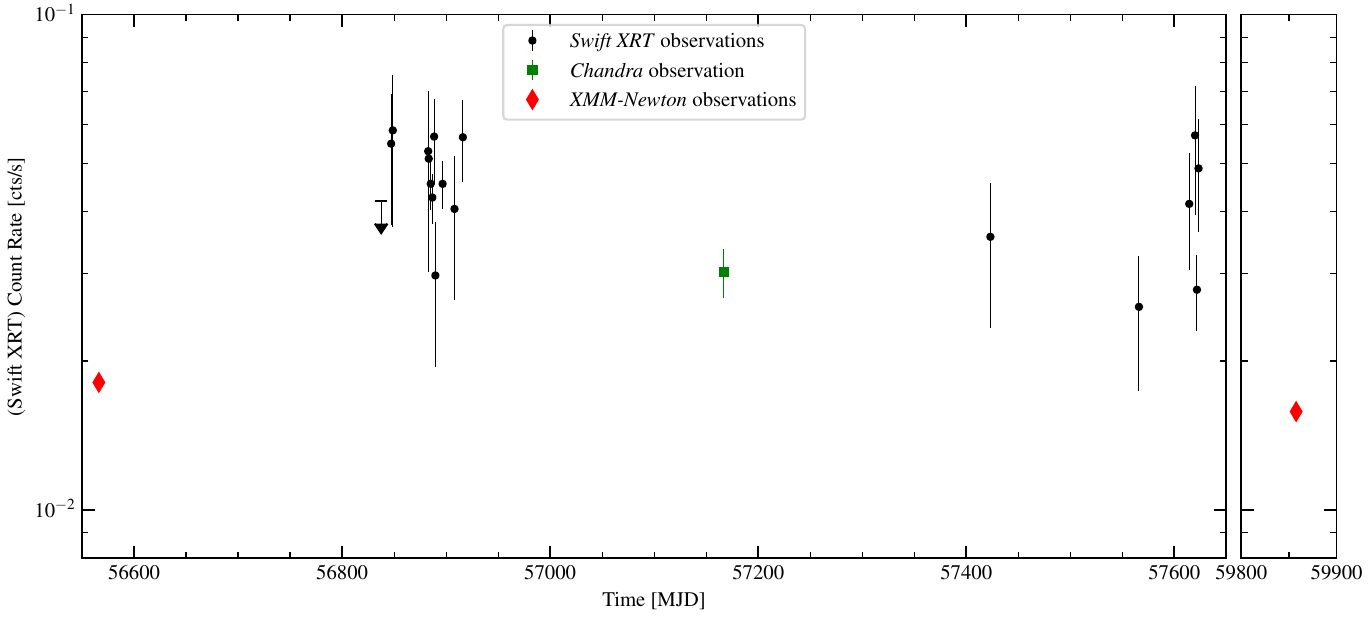}
  \caption{X-ray light curve of IC5052 ULX. Count rates are shown for the $0.3-10~\keV$ band with $1\sigma$ uncertainties. The \chandra (square marker) and \xmm (diamond marker) observations are overplotted after conversion to equivalent \swift XRT count rates. There is minimal long-term variability across the $\sim 9$-year monitoring baseline.}
  \label{fig:obs-countrate}
\end{figure*}

We found the flux of the best-fit model (absorbed powerlaw) of the 2013 \xmm dataset and the \chandra dataset to be a factor of $\approx 1.14$ higher and $\approx 1.9$ higher, respectively, than the model observed flux for the best-fit of the 2022 \xmm dataset. The confidence interval of the observed fluxes for each dataset were obtained by simulating 10,000 mock spectra from the fitted models, using the \texttt{sample\_flux} routine from \texttt{Sherpa}, and taking the flux values for the 16th and 86th weighted quantiles. These flux values and the best-fit model flux value (from Table~\ref{tab:best-fit-model}) were passed through \texttt{WebPIMMS} to obtain the count-rate with the $1\sigma$ confidence interval. From the resulting long-term light curve of Figure~\ref{fig:obs-countrate}, the source appears to be persistent with variations of a factor of $\sim 2-3$ in amplitude, similar to the long-term variability in other persistent ULXs \citep[see][]{pintore2014, fabrika2021}, e.g. NGC5408 X-1 \citep[factor of $\sim 2-3$;][]{grise2013}, IC342 X-1 \citep[factor of $\approx 4$;][]{shidatsu2017}, NGC 1313 X-1 \citep[factor $>3$ at $\sim 3~\keV$;][]{walton2020a}.

\subsection{Short-Timescale Variability Analysis}
\label{sec:variability-analysis}
Accreting black holes in both Galactic X-ray binaries (GXRB) and active galactic nuclei (AGN) typically exhibit low-frequency variability in the form of power-law red-noise or band-limited noise \citep{heil2009, demarco2023}. To assess the short-term variability of IC5052 ULX, we analysed the (fractional rms squared) power spectral density (PSD) from background-subtracted Epn light curves obtained from both the 2013 and 2022 \xmm observations. Instead of using the \texttt{maxSNR} filtering method for the 2022 observation, we filtered with the standard (and more conservative) $0.4~\mathrm{ct s}^{-1}$ to ensure the PSD of the background region light curve was consistent with the predicted Poisson noise.

\begin{figure*}
  \includegraphics[width=\textwidth]{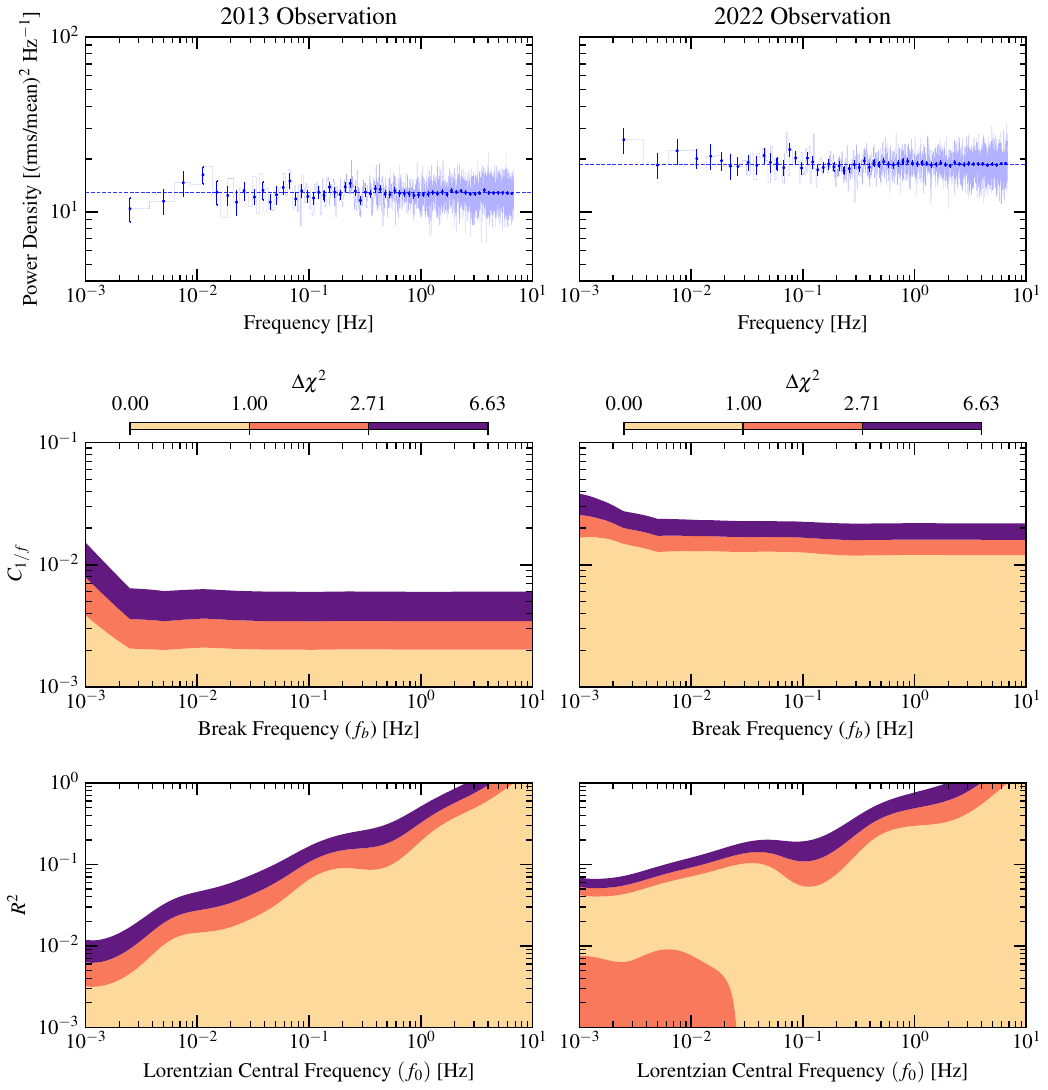}
  \caption{The panels from top to bottom show the averaged power density spectrum, and $\chi^2$ contours for fits to the broken power-law model and to the band-limited noise model, respectively. The left and right column of panels are obtained from the 2013 and 2022 \xmm observation, respectively. In the top panel, the blue points show the (fractional rms squared) power after re-binning the averaged power spectrum by a factor $1.1$ in frequency, while the grey solid line shows the averaged power spectrum without frequency re-binning. The horizontal, grey dashed line shows the Poisson noise floor as estimated from the original source region mean count rate and (area scaled) background region mean count rate. In the bottom two panels, the shaded regions show where the $\Delta \chi^2$ is within the minimal $\chi^2$ for the particular characteristic frequency ($f_b$ or $f_0$).}
  \label{fig:psd}
\end{figure*}

The PSD of the source was estimated from averaging periodograms ($> 20$ segments to enable the use of $\chi^2$-statistics) of $400~\mathrm{s}$ light curve segments. We chose this segment interval to balance low-frequency coverage (probing down to $\sim 2.5~\mathrm{mHz}$) with maximising the available live time in both datasets, yielding $41$ segments from both the 2013 and 2022 observations. We logarithmically re-binned the PSD in frequency by a factor of 1.1 per bin in order to reduce the variance at high frequencies. The top panels of Figure~\ref{fig:psd} show the resulting PSDs, which both appear flat across the whole frequency range and consistent with the predicted Poisson noise level, given by $2 S / (S - B)^2$, where $S$ is the mean (gross) count rate in the source region, $B$ is the area adjusted mean count rate from the background region \citep{vaughan2003}. No excess power is detected above this level, so a direct integration of the Poisson-subtracted PSD (as done by e.g. \citealt{pintore2014}) is consistent with zero fractional variability within the uncertainties. We therefore place upper limits on any intrinsic variability by fitting noise models to the PSDs.

We first tested the fit of a constant model. Both PSDs were well fitted, yielding $\redchisq \sim 56.8 / 58$ and $\redchisq \sim 39.90 / 58$ for the 2013 and the 2022 PSD respectively, confirming the consistency with white-noise variability.

For other noise models, we estimated the Poisson noise level in the PSD by fitting a constant model only to the high-frequency tail, since it is typically unaffected by (low-frequency) red-noise variability \citep{heil2009}. For the PSD of both the 2013 and 2022 \xmm observations, the high-frequency ($f > 0.5~\mathrm{Hz}$) tail of the PSD was found to be in good agreement with the predicted Poisson noise to within the statistical uncertainty of the fit, confirming that the white-noise floor is well calibrated and that no significant excess power is present.

Following \cite{heil2009}, we fitted the PSDs with two different noise models that are typically used to characterise the power spectra of GXRB and AGN: a broken power-law model and a band-limited noise model, to obtain $90\%$ upper-limits to the fractional rms under each scenario. Since the PSDs are actually consistent with Poisson noise across the full frequency range probed, we computed the $\chi^2$-statistic over a range of parameter values, and found upper-limits from $\Delta \chi^2$ contours (assuming 1 d.o.f.).

The broken power-law model is representative of red-noise (flicker) variability, and is given by
\begin{align}
P(f) = \begin{cases}
	C_{1/f} \ f^{-1} + C_P \ &(f < f_b) \\
	C_{1/f} \ f_b f^{-2} + C_P \ &(f \geq f_b)
	\end{cases},
\label{eq:bpl}
\end{align}
where $C_{1/f}$ is the characteristic $1/f$ amplitude, $f_b$ is the break frequency, and $C_P$ is the Poisson noise level (see \citealt{heil2009}). For comparison with \cite{middleton2015}, the fractional rms squared can be obtained from the characteristic amplitude according to
\begin{align}
F_\mathrm{var}^2 &\sim \int_{f_\mathrm{min}}^{f_\mathrm{max}} P(f) - C_P ~ \mathrm{d}f \nonumber \\
&= C_{1/f} \left[ 1 + \log\left( f_b / f_\mathrm{min} \right) - \left( f_b / f_\mathrm{max}\right)\right], \label{eq:bpl-Fvar}
\end{align}
where $f_\mathrm{min}\leq f_b \leq f_\mathrm{max}$. The results of the $\chi^2$-statistic computation, for the range $f_b$ over $10^{-3}-10~\mathrm{Hz}$ and $C_{1/f}$ over $10^{-3}-10^{-1}$, are shown in the second row of panels in Figure~\ref{fig:psd}. Since the PSD is consistent with Poisson noise across the full band, the break frequency is unconstrained by the data. For consistency with the sample of \cite{heil2009}, we evaluate the variability upper limit at break-frequency $f_b = 1~\mathrm{Hz}$ for a frequency band $f_\mathrm{min} = 2.5~\mathrm{mHz}$ to $f_\mathrm{max} = 1~\mathrm{Hz}$ (the band containing most of the expected variance, \citealt{middleton2015}). The $90\%$ upper-limit in the $0.3-10~\keV$ energy range from equation~(\ref{eq:bpl-Fvar}) is $F_\mathrm{var} < 15\%$ for the 2013 \xmm observation and $F_\mathrm{var} < 31 \%$ for the 2022 \xmm observation.

The band-limited noise model is given by a Lorentzian:
\begin{align}
P(f) = 2 R^2 \frac{Q f_0 / \pi}{f_0^2 + Q^2 (f_0 - f)^2} + C_P, \nonumber
\end{align}
where again $C_P$ is the Poisson noise level, $Q$ is the quality factor (with 0.5 used by \citealt{heil2009} as a fiducial value for their band-limited noise model), $f_0$ is the Lorentzian central frequency, and $R^2$ is the normalisation satisfying $\int_0^{\infty} P(f) \mathrm{d}f = R^2$ (c.f. \citealt{heil2009}), which gives the integrated fractional rms squared. The results of the $\chi^2$-statistic computation, for the range $f_0$ over $10^{-3}-10~\mathrm{Hz}$ and $R^2$ over $10^{-3}-1$, are shown in the third row of panels in Figure~\ref{fig:psd}. Choosing a Lorentzian central frequency $f_0 = 10^{-2}~\mathrm{Hz}$ yields the $90\%$ upper-limit in the $0.3-10~\keV$ energy band as $F_\mathrm{var} < 16\%$ for the 2013 \xmm observation and $F_\mathrm{var} < 30\%$ for the 2022 \xmm observation.

Despite the longer total exposure time in the 2022 \xmm observation, the 2013 \xmm observation provides tighter upper-limits to the fractional variability, which can be seen in Figure~\ref{fig:psd} as systematically larger amplitudes consistent with the $90\%$ confidence interval from the minimal $\chi^2$, i.e. $\Delta \chi^2 < 2.71$ for $\nu = 1$. This can be attributed to the lower overall flux in the 2022 \xmm observation (see Figure~\ref{fig:obs-countrate}), which results in a higher Poisson noise floor (visible in Figure~\ref{fig:psd}).

\subsection{Pulsation Search}
\label{sec:pulsation-search}
We searched for evidence of pulsations in the 2022 Epn dataset (without filtering the exposure for background flaring), as well as the joint (light curve summed) 2022 \nustar FPMA + FPMB dataset, using the accelerated search \citep{ransom2002} routines from the \texttt{Stingray} library \citep{huppenkothen2019, bachetti2024}. For the Epn data, we searched both the $0.3-10~\keV$ \xmm energy bandpass and a harder $1.0-10~\keV$ energy band --- the latter motivated by the increase in pulsed fraction with energy observed in ULXPs (e.g. \citealt{bachetti2014, furst2016}). To potentially mitigate the effect of a large frequency derivative typical of ULXPs, we also searched $\approx 29~\mathrm{ks}$ segments of the Epn dataset. We identified any candidate frequencies and frequency derivatives, $(f, \dot{f})$-pair, in the range $0.001-6.812 \ \mathrm{Hz}$, where the upper frequency limit was chosen to be the Nyquist frequency of the \xmm Epn data (in full-frame mode), allowing for a more relaxed $85\%$ significance threshold in this initial stage (for any candidates with more complex than purely sinusoidal profiles) before follow-up significance testing.

Each candidate $(f, \dot{f})$-pair above the relaxed search threshold was further tested using the more sensitive epoch-folding (binned) $Z_n^2$-statistic with $n=2$ harmonics and $12n$ phase bins. We searched a $200\times200$ $(f, \dot{f})$ grid around each candidate, oversampling the independent $f$ and $\dot{f}$ by a factor of $10$. We used $n=2$ harmonics as this provides good sensitivity to both simple sinusoidal profiles and those with some harmonic structure \citep{buccheri1983}, which is the case for the pulse profiles observed in known ULXPs \citep{bachetti2014, furst2016, israel2017, carpano2018, rodriguezcastillo2020}. None of the initial candidate pairs reached $1\sigma$ significance when accounting for the ``look elsewhere" effect, i.e. taking, as a conservative estimate, the independent number of trials from the accelerated search grid rather than the local candidate grid.

Since there were no clear pulsations detected, we calculated the upper-limit to the pulse fraction any signal (with a single harmonic) intrinsically present could have and still remain below the $3\sigma$ detection threshold. We simulated photon arrival-time data using the \texttt{simulate\_times} routine from \texttt{Stingray}, generating 1000 synthetic lightcurves for each pulsed-fraction from $5\%$ to $25\%$ (in steps of $1\%$) for the \xmm data and $>25\%$ for the \nustar data. For each lightcurve, the signal frequency $f$ was uniformly sampled from an interval between $0.1 - 5 \ \mathrm{Hz}$ and the frequency deriative $\dot{f}$ was log-uniformly sampled from $3.06\times10^{-11} - 1.53\times10^{-7}~\mathrm{Hz~s}^{-1}$ with random sign (i.e. $\pm\dot{f} T_\mathrm{obs}^2$ from $0.1$ to $500$, where $T_\mathrm{obs}$ is the total observation duration). The synthetic lightcurves were filtered with the same good-time-intervals as the observations (in the case of the \textit{NuSTAR}, we used the joint FPMA + FPMB) in order to maintain identical windowing effects as in the observations. We then computed the $Z_n^2$-statistic for $n=2$ harmonics using $12n$ phase bins, matching the search procedure of the real data, at the injected $(f, \dot{f})$ pair. The upper-limit pulsed-fraction was taken as as the minimum value at which $90\%$ of the simulated signals produced a $Z_n^2$-statistic exceeding the $3\sigma$ detection threshold. For the \xmm Epn dataset with an energy bandpass of $0.3-10~\keV$, we found an upper-limit pulsed fraction of $\sim 16\%$, and for the \nustar dataset (combining the photons from FPMA and FPMB) with an energy bandpass of $3-25~\keV$, we found an upper-limit pulsed-fraction of $\sim 50\%$.

\section{Optical Counterpart}
\label{sec:optical}
The local region of the edge-on galaxy IC5052 that contains the ULX has been observed on two occasions by the \textit{Hubble Space Telescope (HST)} with the Advanced Camera for Surveys (ACS). For each of the pointings, the same F606W and F814W filter combination was used. Details of the two observations are shown in Table~\ref{tab:hst_observations}. The ULX has also been observed once by \textit{Chandra} with sufficient exposure time to achieve $<0.4$ arcsecond precision in absolute astrometry, registered to the International Celestial Reference System (ICRS) using \textit{Gaia} eDR3 \citep{gaia-collaboration2021}, in the pipeline processing for the \textit{Chandra} Source Catalog (CSC) v2.1 \citep{Evans2024}. The latest \hstacs images available for IC5052 are also registered to the ICRS using \textit{Gaia} eDR3 \citep{mack2022}, allowing us to directly search for optical counterparts without having to perform any additional alignment.

\begin{table}
  \begin{center}
  \caption{Details on the \textit{HST} observations of IC5052.}
  \label{tab:hst_observations}
  \begin{tabular}{lccc}
    \hline
    \hline
    Prop ID & Obs Date & Filter & Exposure Length (s)\\
    \hline
    $9765$ & 2003-12-14 & F606W & 676\\
    $9765$ & 2003-12-14 & F814W & 700\\
    $12196$ & 2011-05-19 & F606W & 2598\\
    $12196$ & 2011-05-19 & F814W & 2599\\
    \hline
  \end{tabular}
  \end{center}
  Data products were obtained from the MAST Archive\footnotemark{}, with search parameters for IC5052 and filtered to only HAP-SVM.
\end{table}
\footnotetext{\url{https://mast.stsci.edu/portal/Mashup/Clients/Mast/Portal.html}}

\begin{figure}
  \includegraphics[width=\columnwidth]{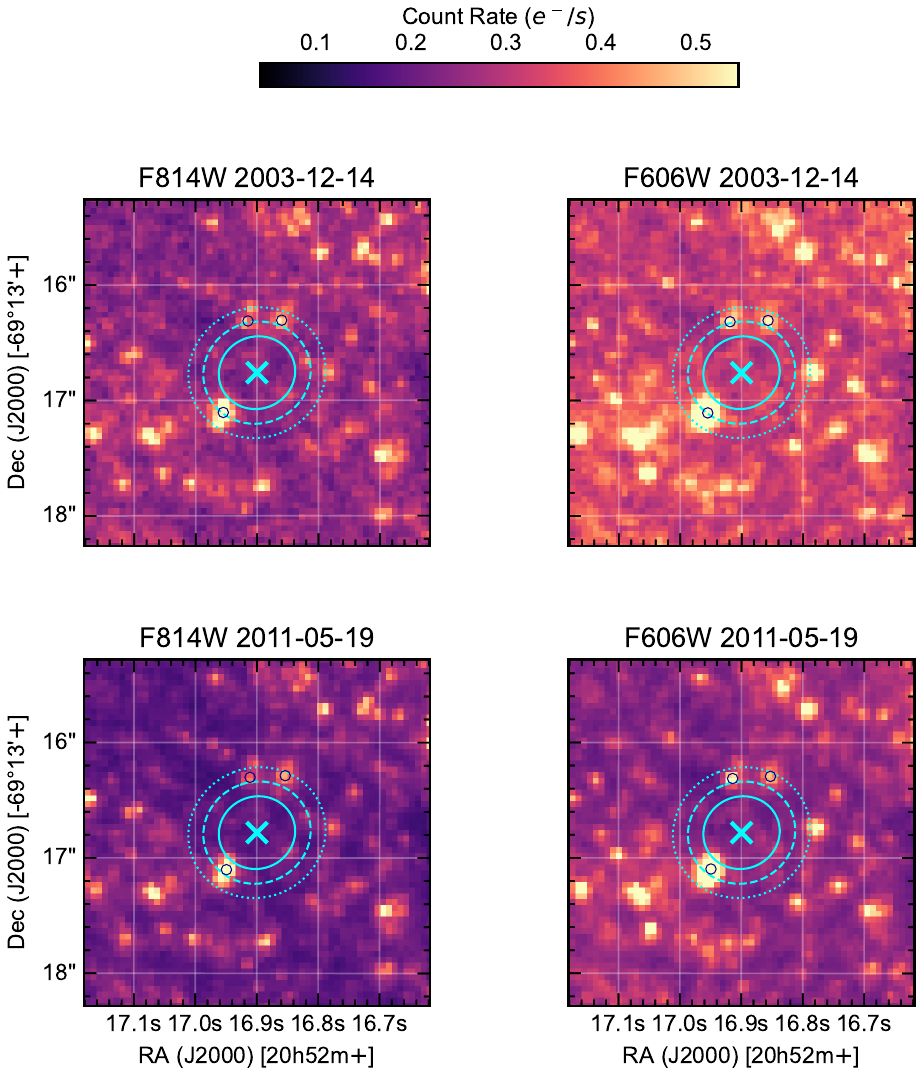}
  \caption{\textit{HST}/ACS observations showing a $60\times60$-pixel region centred on the ULX position from CSC v2.1. For improved contrast of bright sources with the diffuse emission, the pixel intensity range is clipped and rescaled to 1\%-99\% values (linearly). The cyan $\times$ marker indicates the ULX centroid position based on the \chandra data. The solid, dashed, and dotted cyan ellipses show the 95\% (from CSC v2.1), $3\sigma$, and $4\sigma$ uncertainty regions, respectively, for the absolute astrometric position of the ULX. The dark blue $\circ$ markers show the three closest point sources (to the ULX centroid position) from astrometry with \texttt{DAOStarFinder} using $3\sigma$ threshold above the background.}
    \label{fig:hst_obs}
\end{figure}

The panels of Figure~\ref{fig:hst_obs} show a $60 \times 60$-pixel close-up of the \textit{HST}/ACS images centred on the ULX position with the 95\% uncertainty region (obtained from the CSC v2.1), and (rescaled) ellipses showing the $3\sigma$, and $4\sigma$ uncertainty regions overlayed. We searched for candidate optical counterparts first in the accompanying pipeline point-source catalog of each image, filtering for sources within the $60 \times 60$-pixel region of interest. Across all four images, one bright source was coincident with the ULX centroid within $3\sigma$. Two other sources were found with centroid separation just beyond $3\sigma$ (but comfortably within $4\sigma$) in the 2011 observations and are visually discernible from the background in Figure~\ref{fig:hst_obs}, despite not meeting the $5\sigma$ threshold adopted by the pipeline point-source catalog in the 2003 \textit{HST} observations.

We also conducted custom source detection using the \texttt{DAOStarFinder} algorithm from \texttt{photutils} package \citep{bradley2025}, which implements the \texttt{DAOPHOT} star-finding algorithm \citep{stetson1987}. The centroid positions found by this method are shown in Figure~\ref{fig:hst_obs} by the dark $\circ$ markers. For each image, we modelled the diffuse background using 2D median estimation in $27\times27$-pixel boxes with $3\times3$-pixel filtering and sigma-clipping ($3\sigma$, $15$ iterations). From the background-subtracted images, we identified sources at $3\sigma$ above the residual noise level with an assumed point-spread-function full-width-at-half-maximum of $2.8$ pixels for F606W, and $3.0$ pixels for F814W, as recommended \citep{hathi2024}. We applied the standard quality cuts (sharpness $0.2-1.0$, roundness $-1.0-1.0$) to reject contaminants. The positions of the detected sources were cross-matched with a $0.05"$ tolerance between filters and then observational epochs. This analysis recovered all the point-sources from the pipeline catalog, as well as the two sources near the ULX centroid that fell below the pipeline $5\sigma$ threshold in the shorter 2003 exposures.

We measured magnitudes of each of the detected sources using the aperture photometry tools made available in the \texttt{photutils.aperture} sub-package, following the Drizzlepac documentation \citep{goldman2025} for the pipeline point-source catalog. As the candidate counterparts lie within a crowded region of the field, we used a small $3$-pixel aperture for the source measurement. Following the pipeline point-source catalog, we estimated the background for each source from a $0.25"-0.5"$ annulus aperture, but we increased the number of maximum iterations for the sigma-clipped statistics to ensure exclusion of any contaminating bright pixels from nearby sources. We verified our photometry by comparison of position-matched sources in the pipeline point-source catalogs. To calculate the aperture-corrected AB magnitudes, we followed the recommended procedure \citep{hathi2024}: in each exposure, we calibrated the encircled energy fraction for the $3$-pixel aperture to the standard $0.5"$ aperture using isolated, high signal-to-noise ratio sources in the field, and then using the tabulated fractions \citep[Table~5.1]{hathi2024} for the correction from $0.5"$ to infinite aperture. For each of the candidate counterparts, we found the magnitudes for each filter to be consistent (within $1\sigma$ uncertainty overlap) across epochs and so calculate a final set of weighted average magnitudes combining the data from both epochs.

The measured magnitudes were corrected for Galactic dust absorption along the line of sight to the ULX position using the dust maps compiled by \citet{schlegel1998} and recalibrated by \citet{schlafly2011}. Extinction coefficients for the \textit{HST} ACS F606W and F814W filters were obtained from the NASA/IPAC Extragalactic Database (NED)\footnote{\url{https://ned.ipac.caltech.edu/}}, giving $A_{\mathrm{F606W}} = 0.125 \ \mathrm{mag}$ and $A_{\mathrm{F814W}} = 0.077 \ \mathrm{mag}$. The dereddened apparent magnitudes and $m_\mathrm{F606W}-m_\mathrm{F814W}$ colours for each candidate counterpart are tabulated in Table~\ref{tab:counterpart_properties}. We calculated absolute magnitudes assuming a distance modulus to IC5052 of $(m-M)_0=28.70\pm0.10$ \citep{tully2013}.

\begin{table*}
  \begin{center}
  \caption{Properties of each candidate counterpart.}
  \label{tab:counterpart_properties}
  \begin{tabular}{ccccccccc}
    \hline
    \hline
    ID & RA & DEC & Separation [$\sigma$] & $m_\mathrm{F606W}$ & $m_\mathrm{F814W}$ & $C^{\dagger}$ & $M_\mathrm{F606W}$ & $M_\mathrm{F814W}$ \\
    \hline
    1 & 20:52:16.95 & $-$69:13:17.1 & $2.67$ & $23.24\pm0.01$ & $23.43\pm0.02$ & $-0.14\pm0.03$ & $-5.33\pm0.10$ & $-5.19\pm0.10$\\
    2 & 20:52:16.91 & $-$69:13:16.3 & $3.22$ & $25.00\pm0.07$ & $24.73\pm0.08$ & $~~~0.32\pm0.11$ & $-3.57\pm0.12$ & $-3.90\pm0.13$\\
    3 & 20:52:16.85 & $-$69:13:16.3 & $3.53$ & $24.95\pm0.06$ & $24.25\pm0.05$ & $~~~0.75\pm0.08$ & $-3.63\pm0.12$ & $-4.38\pm0.11$\\
    \hline
  \end{tabular}
  \end{center}
The separation is given in terms of the number of $\sigma$ deviations from the ULX centroid position. \\
Dereddened magnitudes, $m$ and $M$, as well as the colours, $C$, are given in the AB system. \\
$^\dagger$: The colour is calculated from $m_\mathrm{F606W} - m_\mathrm{F814W}$. \\
The absolute magnitudes, $M$, are calculated from the apparent magnitudes, $m$, using a distance modulus of $28.70\pm0.10$.
\end{table*}

\begin{figure}
  \includegraphics[width=\columnwidth]{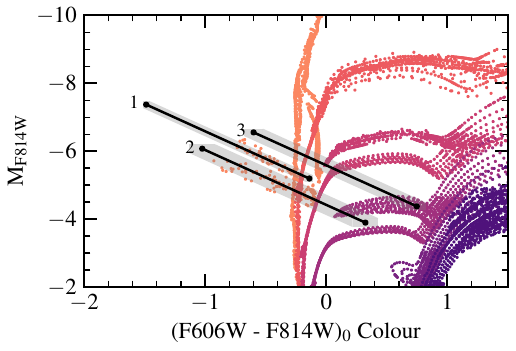}
  \caption{Intrinsic colour-magnitude diagram for stellar populations simulated using \texttt{PARSEC}. Marker colours progress from orange to purple with increasing stellar age, from $\sim 3.2~\mathrm{Myr}$ to $\sim 1000~\mathrm{Myr}$; the spread within each colour reflects the range of metallicities modelled, from $Z \sim 0.004$ to $\sim 0.016$. The dereddened colour-magnitude positions of the three candidate counterparts are overplotted in black, with numbered labels corresponding to the counterpart IDs in Table~\ref{tab:counterpart_properties}. For each candidate, the two extrema connected by a line represent the range of possible intrinsic positions: the left endpoint assumes the maximum reddening implied by the X-ray derived $N_\mathrm{H}$, while the right endpoint assumes only Galactic extinction along the line of sight. The grey shaded regions indicated photometric uncertainties.}
  \label{fig:isochrones}
\end{figure}

To estimate the stellar properties of each of the candidate counterparts, we used \texttt{PARSEC} isochrones \citep{bressan2012}, which provide stellar evolutionary tracks with extensive coverage of post-main-sequence phases and a wide range of metallicities, configured for \textit{HST} ACS/WFC passbands with 2021 zero-point corrections. These were accessed through the \texttt{ezpadova} package \citep{Fouesneau_ezpadova_2025}. Since there are no measurements of the metallicity of IC5052 that we are aware of, we used a wide range of initial metallicities from $Z = 0.004$ to above solar-like $Z = 0.016$, and matched each candidate counterpart by $\chi^2$-minimisation for dereddened colour and magnitude. We tested a range of extinction coefficients (in addition to the Galactic dust extinction correction, which is always applied) from $A_\mathrm{V} = 0$ to $\sim 3.9$ mag, which corresponds to a hydrogen column density of $N_\mathrm{H} \approx 0.7 \times 10^{22} \ \mathrm{cm}^{-2}$ (slightly above the best-fit $N_\mathrm{H}$ from our X-ray spectral modelling; see Table~\ref{tab:best-fit-model}), using the conversion $N_\mathrm{H} / A_\mathrm{V} = 1.79 \times 10^{21} \ \mathrm{cm}^{-2} / \mathrm{mag}$ \citep{predehl1995} as previously done by \cite{roberts2008a}. The colour-magnitudes of each of the candidate counterparts for the various (additional X-ray) extinction coefficients is plotted in black over the \texttt{PARSEC} isochrones in Figure~\ref{fig:isochrones}.

For larger extinction coefficients approaching the X-ray spectrum best-fit hydrogen column density (shown as the left-most extrema of the lines in Figure~\ref{fig:isochrones}), the colour of the brightest candidate source (shown in Figure~\ref{fig:isochrones} with the most negative set of $\mathrm{M}_{\mathrm{F814W}}$) becomes implausibly blue compared to all isochrones. More generally, increasing the extinction coefficient systematically shifts the inferred stellar parameters toward more massive, earlier spectral types for all three candidate counterparts.

The brightest candidate counterpart (labelled with ID 1 in Table~\ref{tab:counterpart_properties}) shows the most consistent match across the range of extinction values. For low extinction ($A_\mathrm{V} \lesssim 0.35$), it most closely matches a subgiant with an age of $\sim 10~\mathrm{Myr}$ and mass $\sim 19~\solarM$. For higher extinction ($A_\mathrm{V} \gtrsim 0.35$), the source tracks along the $\sim 3.2~\mathrm{Myr}$ isochrone (shown by the orange points in Figure~\ref{fig:isochrones}), with the inferred mass increasing from $\sim 39~\solarM$ to $\sim 49~\solarM$ at the highest extinctions. Both of these are consistent with high mass X-ray binary (HMXB) donors observed in other systems.

The two fainter candidate counterparts (labelled with ID 2 and 3 in Table~\ref{tab:counterpart_properties}) are more difficult to constrain, as they intersect a broader range of isochrone tracks depending on the assumed extinction. For low extinction ($A_\mathrm{V} \lesssim 1$), they match stellar populations with masses $\sim 5~\solarM$. For intermediate to high extinction ($A_\mathrm{V} \sim 1$ to $2$), the best-fit masses range from $\sim 8~\solarM$ to $\sim 18~\solarM$.

\section{Discussion}
\label{sec:discussion}
\subsection{Broadband X-ray Continuum Emission}
\label{sec:broadband-continuum-emission}
For the sample of ULXs with simultaneous X-ray broadband observations, and where the \nustar detection has been up to at least $20~\keV$, their spectra have systematically required a third spectral continuum component to capture the otherwise hard excess \citep{walton2018}. Unlike the rest of the broadband sample, the broadband dataset for IC5052 ULX shows a sufficiently good fit to a double thermal-component model, but with an implausibly high inner disc temperature of $\Tin \approx 6.4~\keV$ - well outside of the range $\sim 1 - 4~\keV$ seen from other ULXs in the broadband sample (see Table 3 of \citealt{walton2018}). In contrast, both of the three-component models that have typically been used to describe the broadband spectra of ULXs in recent works, the magnetic accretor model (\S\ref{sec:magacc}) and the non-magnetic accretor model (\S\ref{sec:bhmodel}), allow for a parameter space that is broad enough to overlap with the range of temperatures seen from the rest of the broadband ULX sample, while achieving comparable goodness-of-fit to the two-component thermal model.

Using the spectral classification criteria (of \citealt{sutton2013}), the \xmm only spectrum places IC5052 ULX within the Hard Ultra-Luminous State (HUL; see the final paragraph of \S\ref{sec:spectroscopy}). This HUL state is generally thought to be associated with a face-on viewing orientation, where we observe down the evacuated funnel of a super-Eddington accretion flow \citep{shakura1973, poutanen2007, middleton2015}. The viewing geometry is crucial for interpreting the origin of the spectral components. However, these spectral states were distinguished using only \xmm and \chandra observations, which have a more limited X-ray bandpass compared to the simultaneous broadband \nustar observations.

The high ($\Tin \approx 6.4~\keV$) inner disc temperature of the double thermal-component model cannot simply be understood as an exceptional case in the broadband ULX sample. Instead, it implies there is a problem with interpreting the hard emission ($> 10~\keV$) as originating unmodified from a super-Eddington disc. Comparing to calculations of the super-Eddington disc of \cite{poutanen2007}, which includes advection and outflows, the maximum effective temperature for a super-Eddington flow is approximately given by $T \approx 1.6~m^{-1/4}~\keV$, where $m$ is the compact object mass in units of $\solarM$. Even allowing for a large colour-correction factor, $f_\mathrm{col} \sim 3$, the fitted temperature would require an implausibly small compact-object mass.

The implausibility of the double thermal-component best-fit is also evident from the implied characteristic size of the emitting region. The apparent inner radius of a multi-colour disc component is related to its normalisation parameter, $N$, by $R_\mathrm{in} = \xi f_\mathrm{col}^2 (N / \cos i)^{1/2} D_{10}~\mathrm{km},$ where $D_{10}$ is the distance in units of $10~\mathrm{kpc}$ ($D_{10} = 550$ at $5.5~\mathrm{Mpc}$), $i$ is the inclination angle of the line-of-sight to the disc, and $\xi$ is a (geometric) factor correcting for the displacement of the peak temperature from the inner disc boundary. For the \texttt{diskpbb} component, the fitted value $p\approx 0.5$ is consistent with an advection-dominated temperature profile. Adopting the corresponding boundary correction factor $\xi = 0.353$ \citep{vierdayanti2008} and colour correction factor $f_\mathrm{col} = 3$ \citep[e.g.][]{soria2015, brightman2016}, the best-fit normalisation $N \approx 9.0 \times 10^{-6}$ (Table~\ref{tab:best-fit-model}) yields $R_\mathrm{in} \lesssim 6.2~\mathrm{km}$ for $\cos i > 0.7$ ($i < 45^\circ$), which is implied by the HUL classification. For the inner radius (without beaming) to match the innermost stable circular orbit of a black hole requires an unphysically small Schwarzschild black hole of mass $\lesssim 0.5~M_\odot$, or a maximally spinning (astrophysically limited $a_* = 0.998$; \citealt{thorne1974}) Kerr black hole of mass $\lesssim 3.5~M_\odot$. These limits confine any maximally-spinning black hole to a narrow mass range, $2.2 \lesssim m \lesssim 3.5$, where the lower bound is approximately the observationally inferred upper end of the neutron star mass distribution \citep{alsing2018}. However, this window is closed entirely once mild beaming, which is expected in the super-critical picture ($1/b \gtrsim 2.5$; \citealt{poutanen2007, king2023}), is included: the beaming-corrected radius $R_\mathrm{in;\,true} = \sqrt{b}\,R_\mathrm{in;\, apparent}$ implies a maximally-spinning Kerr black hole below the maximum neutron star mass, leaving no self-consistent black hole solution.

\subsubsection{Non-Magnetic Accretor Model}
Compton up-scattering (modelled here by convolution with a \texttt{simpl} component) can account for the hard emission without requiring an exceptionally high temperature from the \texttt{diskpbb} component, particularly if the colour-correction factor for super-Eddington discs exceeds the typically assumed $f_\mathrm{col} \approx 1.7$. However, an up-scattering component introduces a different tension with the standard picture of super-Eddington accretion, in which the funnel is assumed to be evacuated or optically thin \citep{poutanen2007, middleton2015}. For IC5052 ULX, more plausible disc temperatures ($\Tin \sim 3.0~\keV$, compared to the $\approx 6.4~\keV$ for the double thermal-component fit) are favoured only when the up-scattering fraction is substantial, indicating a high degree of reprocessing of the disc emission. This requires corona-like material that is both sufficiently hot (to up-scatter rather than down-scatter the relatively hot inner disc $\Tin \sim 3.0~\keV$) and with sufficient scattering opacity (to reprocess a substantial fraction of disc photons). Radiation magneto-hydrodynamical (RMHD) simulations of super-Eddington accretion onto stellar-mass black holes show such conditions can occur \citep{jiang2014}. \cite{mills2024} post-processed 3D RMHD simulations with Monte Carlo radiative transfer and found that the hot Comptonizing gas in the funnel regions above the inner disc produces hard X-ray emission consistent with that observed in the broadband \xmm and \nustar data of NGC~1313~X-1. In these simulations, the corona is not a distinct structure within an otherwise evacuated funnel (requiring its own stability), but is the hot atmosphere of the super-critical flow itself.

The corona-like interpretation (whether more local to the compact object or extended throughout) can be examined in light of the source's short-term variability constraints (see \S\ref{sec:variability-analysis}). In GXRBs and AGNs, the high-energy power-law component from the corona is typically highly variable \citep{demarco2023}. Similarly, \cite{walton2014} found tentative evidence for higher short-timescale variability in Holmberg IX X-1 in the $10-30~\keV$ band ($F_\mathrm{var} \sim 13\%$) where the Comptonizing component becomes prominent, compared to lower variability ($F_\mathrm{var} \sim 7\%$) in the $3-10~\keV$ band, where the thermal disc emission dominates. A more prominent Comptonizing component, as favoured for fits with lower inner disc temperature (\S\ref{sec:bhmodel}), might therefore be expected to produce some short-term variability.

In the case of IC5052 ULX, the lack of observed variability (see \S\ref{sec:variability-analysis}) could therefore appear to be in tension with a prominent Comptonizing component, which should still contribute significant photon flux (e.g. $\approx 42\%$ in the saturated limit) in the $0.3-10~\keV$ band. However, the $F_\mathrm{var}$ upper limits cannot exclude low intrinsic variability similar to what is observed in Holmberg IX X-1, and any apparent tension between the timing and spectroscopic analyses is further relaxed when incorporating spectral information into the variability constraints. Once dilution from less-variable thermal components is accounted for, the upper limits for the intrinsic coronal variability become looser: $\lesssim 35\%$ for the broken power-law noise model and $\lesssim 38\%$ for the band-limited noise model, which is not incompatible with other coronal systems \citep{demarco2023}.

Overall, the lack of low-frequency variability is consistent with a face-on viewing geometry in the standard picture of super-Eddington accretion \citep{middleton2015}. In this scenario, a more face-on line of sight largely avoids dense clumps in the outflowing material that would impart extrinsic variability. However, these spectral-timing models may require additional development in light of the new broadband spectroscopy findings; in the case of IC5052 ULX, the significant contribution from the hard Comptonized component ($\approx 50\%$ of the energy flux in the $0.3-10~\keV$ band) is otherwise unaccounted for in such models. The radiative transfer calculations of \cite{mills2024} provide a possible physical origin for the Comptonizing component, but do not yet extend to predictions of variability.

\subsubsection{Magnetic Accretor Model}
Similarly to the Compton up-scattering model, the magnetic accretor spectral model (see \S\ref{sec:magacc}) can account for the hard emission, and in this case reduces the temperature of the \texttt{diskpbb} component to $\Tin \sim 1.2~\keV$ according to the best-fit (detailed in Table~\ref{tab:best-fit-model}), although the exact value likely depends on the choice of the CPL shape parameters ($\Gamma$, $\text{HighECut}$) for the accretion column component. The model best-fit is in good agreement with the rest of the broadband ULX sample \citep{walton2018}, and falls within the temperature range predicted by the \cite{poutanen2007} model. However, modifications are required to properly account for the magnetic field at the inner disc boundary and inability to advect energy over the event horizon (see \citealt{chashkina2017, chashkina2019} for models of super-Eddington discs around a magnetic accretor).

Initially, the lack of observed pulsations (see \S\ref{sec:pulsation-search}) might appear to contradict the magnetic accretor model, which was originally invoked to explain ULXP spectra \citep{pintore2017}. \cite{walton2018} noted the non-pulsed ULXs fitted with this model tend to have low \texttt{cutoffpl} flux fractions ($F_\mathrm{col} <60\%$ in the $0.3-40~\keV$ band; obtained from spectral fitting with the main \texttt{cutoffpl} parameters fixed to the average values found from isolating the pulsating emission component via phase-resolved analysis of known ULXPs), in contrast to the known ULXPs (with IC342 X-2 being the only exception; see Table~4 by \citealt{walton2018}). In the magnetic accretor model, the \texttt{cutoffpl} component is interpreted as emission from the neutron star accretion column (or close to the surface) and therefore it is responsible for any observed coherent pulsations. A low flux fraction for this component naturally explains non-detection of pulsations: the smaller number of pulsed photons from the accretion column are diluted by non-pulsed disc emission, and pulsation detection becomes unlikely under the typical observational constraints (which result in pulsation detection thresholds $\sim 16\%$ pulsed-fraction). However, IC5052 ULX presents an interesting borderline case of the empirical threshold identified by \cite{walton2018}, since the local best-fit magnetic accretor model (presented in Table~\ref{tab:best-fit-model}) yields $F_\mathrm{col}$ in the range $65.1 - 69.2\%$ (extrapolating to the $0.3-40~\keV$ range for comparison, see \S\ref{sec:magacc}).

The borderline $F_\mathrm{col}$ under the magnetic accretor model motivates a more detailed examination of the expected pulsed-fraction properties, both from a statistical and physical perspective. During some observational epochs of ULXPs, pulsations are not detected (see e.g. \citealt{bachetti2014, israel2017, sathyaprakash2019, rodriguezcastillo2020, bachetti2020, belfiore2024, imbrogno2024}) but the origin (whether physical or largely statistical in nature) of these pulsation non-detections is unknown \citep[and appears not to be related to geometric effects from its orbit in the case of M51 ULX-7, see][]{rodriguezcastillo2020}. A further complication for the interpretation of pulsation non-detection comes from the increase in pulsed fraction with energy, which has been consistently observed in ULXPs (e.g. \citealt{bachetti2014, furst2016}). However, for the pulsed-fraction upper limits computed over the full \xmm energy range ($0.3-10~\keV$), the harder accretion column emission is diluted by the dominant soft thermal disc emission (in the description of the magnetic accretor model). The computed upper limits correspond to an effective pulsed-fraction across the entire energy range. For IC5052 ULX, we can translate between the effective pulsed-fraction and the intrinsic pulse-amplitude by assuming the \texttt{cutoffpl} component represents the spectral shape of the pulsed emission, as has been demonstrated through phase-resolved spectroscopy for confirmed ULXPs with sufficient data quality \citep{brightman2016, walton2018a, walton2018}. Under this assumption, the $\sim 16\%$ effective pulsed-fraction upper limit from the $0.3-10~\keV$ energy band corresponds to a $\sim 40\%$ upper limit on the intrinsic pulsed amplitude of the \texttt{cutoffpl} component itself. Similarly, the $\sim 50\%$ effective pulsed-fraction upper limit from the $3-25~\keV$ energy band corresponds to a $\sim 63\%$ upper limit to the pulsed amplitude. These pulsed amplitude upper limits are consistent with the pulsed-fractions seen from known ULXPs at the highest energies probed by {\it NuSTAR}, where the contamination by the thermal components is minimal (and so the pulsed-fraction values would more readily be interpreted as the intrinsic pulsed-fraction of the emission from the accretion column).

The weaker constraint on the intrinsic pulsed amplitude (compared to the effective pulsed-fraction upper limit obtained from \xmm data) means IC5052 ULX remains a possible magnetic accretor system despite the absence of observed pulsations. We caveat that these pulse-amplitude upper limits are obtained by extrapolation of the average cutoff power-law shape from phase-resolved spectroscopy of other systems, and in order to fully contextualise the pulse-amplitude upper limits, further theoretical development is required, which is beyond the scope of this work.

The lack of low-frequency variability observed for IC5052 ULX is also consistent with the interpretation as a magnetic accretor within the model of an optically thick envelope \citep{mushtukov2019}. In this model, the high accretion rate onto the neutron star itself results in an optically thick flow along the channelling magnetic field lines \citep{mushtukov2017}, consequently reprocessing half of the accretion column emission into a softer multicolour-thermal spectrum \citep{brice2023}. However, under such a model, our interpretation of the pulse-amplitude as a change to the \texttt{cutoffpl} normalisation is complicated. Phase-resolved ray-tracing has shown that coherent pulsations can be attributed to the co-rotating envelope itself, and this origin can be brought to match the rise in pulsed-fraction with energy in the \xmm band \citep{brice2023, conforti2025}. However, these calculations do not yet self consistently include the hard emission directly escaping the accretion column into the phase-resolved spectrum, nor self-consistently include the geometric effects from a super-Eddington accretion disc.

\subsubsection{Comparison with Prior Work}
\cite{cruz-sanchez2025} independently analysed the same broadband 2022 \xmm and \nustar dataset with both the non-magnetic and magnetic accretor models. Unlike in this work, they did not report the double thermal-component model, which we find to be statistically acceptable yet physically implausible (\S\ref{sec:broadband-continuum-emission}) and thus motivating the three-component spectral model used for other ULXs with broadband X-ray coverage \citep{walton2018}. Comparing our best-fit values (Table~\ref{tab:best-fit-model}) with those of \cite{cruz-sanchez2025} reveals a mix of agreement and disagreement. The parameters that agree (to within $90\%$ uncertainties) with our analysis include the intrinsic column density ($N_\mathrm{H} \approx 0.6 \times 10^{22}~\mathrm{cm}^{-2}$ here, versus $0.64 - 0.7 \times 10^{22}~\mathrm{cm}^{-2}$), the cool \texttt{diskbb} temperature ($\Tin \approx 0.27~\keV$, versus $0.25~\keV$ and $0.30~\keV$), and the hot \texttt{diskpbb} temperature ($\Tin \approx 1.2~\keV$, versus their $1.0~\keV$) for the magnetic accretor model. In contrast, our cross-calibration factors between \xmm Epn and \nustar FPMA/FPMB fall well below the $< 10\%$ expected from instrument cross-calibration \citep{madsen2015}, whereas those of \cite{cruz-sanchez2025} show $> 12\%$ discrepancy. Our fitting with the non-magnetic accretor model favours a substantially hotter inner disc ($\Tin \approx 3.0~\keV$) with a saturated scattering fraction ($\mathrm{FracSctr} \sim 1$), whereas they report $\Tin \approx 1.1~\keV$ with a constrained $\mathrm{FracSctr} \approx 0.8$. The radial temperature index likewise differs: our fits favour an advection-dominated disc ($p \approx 0.5$), whereas \cite{cruz-sanchez2025} report $p=0.7\pm0.2$, nearer the standard \cite{shakura1973} thin-disc value ($p=0.75$), although their uncertainty spans the advection-dominated disc value.

We attribute these differences primarily to the distinct approaches used in the spectral fitting. \cite{cruz-sanchez2025} used the standard Monte Carlo Markov Chain (MCMC) method within XSPEC. Our own analysis reveals that the $\chi^2$-surfaces for IC5052 ULX show substantial complexity, including local minima and noisy features that require robust and iterative fitting methods, e.g. iterating the standard optimisation with the \texttt{Simplex} method in \texttt{Sherpa}, to obtain the final clean contours shown in Figure~\ref{fig:contours}. When fitting the non-magnetic accretor model (\S\ref{sec:bhmodel}) starting from $p = 0.75$ (for the \texttt{diskpbb} component) without applying the iterative strategy, we found results that matched more closely with those reported by \cite{cruz-sanchez2025}, albeit at higher $\redchisq$ than our final best-fit reported in Table~\ref{tab:best-fit-model}. The complexity of the $\chi^2$-surface from the broadband dataset challenges the effectiveness of standard MCMC sampling methods for fully characterising the parameter space, motivating instead more robust methods such as nested sampling \citep{buchner2021}, which are better suited to the multi-modal, boundary-dominated posteriors encountered here.

\subsection{Potential Optical Counterparts}
We successfully identified a small set of candidate counterparts by using the improved astrometric precision from the CSC v2.1 (registered to Gaia eDR3) and the Hubble Advanced Products corrected to the same reference frame. Previous counterpart searches for IC5052 ULX were unsuccessful \citep{chatterjee2016}, primarily due to the larger astrometric uncertainties from the earlier constraints on the \chandra position. Given that the brightest candidate counterpart is located within $3\sigma$ positional coincidence, and shows properties consistent with a post-main-sequence HMXB donor similar to those observed in other ULXs (e.g. \citealt{motch2011, heida2019, heida2019a, gladstone2013, fabrika2021}) for a range of extinction factors and metallicities (see \S\ref{sec:optical}), we consider this to be the most likely optical counterpart to IC5052 ULX. However, confirmation requires additional observations to refine the ULX position and spectroscopic follow-up to establish the nature of the counterpart.

An intriguing tension emerges when applying the extinction coefficient derived from the X-ray spectral best-fit hydrogen column density to the \texttt{PARSEC} isochrones: the dereddened colour of the candidate counterpart becomes implausibly blue (see \S\ref{sec:optical}). This would suggest that the majority of the intrinsic X-ray absorption is localised to the immediate environment of the compact object, rather than distributed along the shared line of sight to the candidate counterpart. Such localised absorption is consistent with the picture of outflows from a super-Eddington disc. However, substantial localised absorption may also introduce tension with the HUL spectral classification inferred from the $0.3-10~\keV$ spectrum, which as previously discussed is thought to be associated with unobscured viewing geometry of the central regions near the compact object, down an evacuated funnel of the super-Eddington accretion flow. Instead, the high intrinsic column density suggests that significant material remains present at high latitudes above the disc. Future high-resolution X-ray spectroscopy with instruments optimised for detecting absorption features (e.g. \textit{NewAthena}) would be valuable to distinguish between kinematic signatures of strong outflows versus more static absorbing structures, helping to constrain the geometry of the system.

We caution that there are alternative interpretations of the optical counterpart beyond direct stellar emission. \cite{tao2011} argued that the bright optical counterpart emission of ULXs may be dominated by X-ray reprocessing (or in the case of NGC 2403 X-1, direct emission) by the outer accretion disc rather than stellar emission, based on observed optical spectral slopes ($F_\nu \propto \nu^\alpha$ with $\alpha \sim -1$ to $2$) and X-ray-to-optical flux ratios - specifically $\xi = B_0 + 2.5 \log F_X$, where $B_0$ is the dereddened $B$ magnitude and $F_X$ is the X-ray flux in $2 - 10~\keV$ band given in $\mu\mathrm{Jy}$, resembling those of low-mass X-ray binaries, for which disc irradiation is well-established \citep{vanparadijs1994}. In the absence of high quality spectroscopy, distinguishing between the stellar-dominated and disc-reprocessing scenarios requires either fitting the optical spectral energy distribution or at least computing $\xi$, but the available two-band photometry ($\sim V$ and $I$; F606W and F814W) for IC5052 ULX is insufficient to constrain the spectral shape and without good approximation for Johnson-Cousins $UBVRI$ $B$ bandpass.

\section{Conclusions}
\label{sec:conclusions}
We presented a comprehensive X-ray broadband spectral and timing analysis of IC5052 ULX, supplemented by improved optical counterpart identification.

While a double thermal-component continuum model provides a statistically reasonable fit to the broadband spectrum, it yields an implausibly high inner disc temperature ($\Tin \approx 6.4~\keV$), inconsistent with super-Eddington disc models and demonstrating that statistical adequacy alone is insufficient for model selection. Both three-component spectral models: the magnetic accretor (DBB + DPBB + CPL) and non-magnetic accretor model (DBB + SIMPL$\otimes$DPBB), provide comparable statistical fits with physically reasonable parameters.

Timing analysis initially appears to reveal tensions with both spectral models. The magnetic accretor model places IC5052 ULX at a borderline accretion column flux fraction ($F_\mathrm{col} \approx 62\%)$, near where empirical trends predict detectable pulsations. The non-magnetic accretor model initially appears challenged by a lack of short-term variability ($F_\mathrm{var} < 15\%$ in the $0.3-10~\keV$ band), seemingly inconsistent with highly variable coronae in black hole accretors.

However, incorporating spectral information relaxes the timing constraints for both models. For the magnetic accretor model, translating the more constraining effective pulsed-fraction upper limit to intrinsic pulse-amplitude constraints yields corrected limits of $\sim 40\%$, allowing IC5052 ULX to remain physically plausible as a magnetic accretor given that known ULXPs (NGC1313 X-2) show pulsed fractions as low as $\sim 5\%$. For the non-magnetic accretor model, accounting for spectral dilution from the dominant thermal components yields intrinsic coronal variability upper limits of $\lesssim 35\%$ (broken power-law model), which remains compatible with coronal systems, particularly given potential energy-dependent variability effects and the different physical conditions in super-Eddington versus sub-Eddington accretion. This highlights how spectral dilution effects may need to be incorporated when interpreting pulsation non-detections, with systematic application of such techniques essential for characterising the ULX parameter space and informing future observational strategies.

We identified a promising optical counterpart consistent with an evolved HMXB donor, with discrepant absorption geometry between X-ray and optical constraints that suggests localised X-ray absorption, possibly from super-Eddington disc winds. Future high-resolution X-ray spectroscopy (e.g. by \textit{XRISM} / \textit{NewAthena}) and optical spectroscopy would be valuable for characterising any potential outflows and definitively establishing the counterpart association.

\section*{Acknowledgements}
We thank the anonymous referee for their careful and thorough reading of the manuscript, and for suggestions which have improved the clarity and rigour of this work.

NB and DJW acknowledge support from the Science and Technology Facilities Council (STFC; grant code ST/Y001060/1). TPR and AHK acknowledge support from STFC as part of the consolidated grant award ST/X001075/1.

This research is based on observations obtained with \textit{XMM-Newton}, an ESA science mission with instruments and contributions directly funded by ESA Member States and NASA. This research has also made use of data obtained with \textit{NuSTAR}, a project led by Caltech, funded by NASA and managed by NASA/JPL, and has utilized the NuSTARDAS software package, jointly developed by the ASDC (Italy) and Caltech (USA). This research is additionally based on observations made with the NASA/ESA Hubble Space Telescope, obtained from the Data Archive at the Space Telescope Science Institute, which is operated by the Association of Universities for Research in Astronomy, Inc., under NASA contract NAS 5-26555. These observations are associated with program $\#9765$ and $\#12196$.

This research has made use of the NASA/IPAC Extragalactic Database (NED), which is funded by NASA and operated by Caltech.

%%%%%%%%%%%%%%%%%%%%%%%%%%%%%%%%%%%%%%%%%%%%%%%%%%
\section*{Data Availability}
 
The X-ray data analysed in this work are publicly available from the \xmm Science Archive, \chandra Data Archive, \nustar archive at HEASARC, and \swift Data Archive. Observation IDs are provided in Table~\ref{tab:observations}. Optical data were obtained from the Hubble Legacy Archive (see \S\ref{sec:optical}).



\bibliographystyle{mnras}
\bibliography{../../../bib/bib} % if your bibtex file is called example.bib

%\begin{thebibliography}{99}
%\end{thebibliography}

%%%%%%%%%%%%%%%%%%%%%%%%%%%%%%%%%%%%%%%%%%%%%%%%%%

%%%%%%%%%%%%%%%%% APPENDICES %%%%%%%%%%%%%%%%%%%%%

% \appendix

% \section{Some extra material}

%If you want to present additional material which would interrupt the flow of the main paper, it can be placed in an Appendix which appears after the list of references.

%%%%%%%%%%%%%%%%%%%%%%%%%%%%%%%%%%%%%%%%%%%%%%%%%%

% Don't change these lines
\bsp	% typesetting comment
\label{lastpage}
\end{document}